\documentclass[11pt]{article}
\pdfoutput=1

\usepackage[margin=1in]{geometry}
\usepackage{amsfonts}
\usepackage{amsmath}
\usepackage{graphicx}
\usepackage{algorithm}
\usepackage[noend]{algpseudocode}
\usepackage{url}
\usepackage{multirow}
\usepackage{authblk}
\usepackage{rotating}
\usepackage{xr}

\usepackage[style=nature,maxbibnames=5,mincitenames=1,maxcitenames=1]{biblatex}
\renewbibmacro{in:}{%
  \ifentrytype{article}{}{%
  \printtext{\bibstring{in}\intitlepunct}}}
\newcommand{\degree}{\ensuremath{^\circ}}

\begin{document}

\title{{\em Cloudbreak}: Accurate and Scalable Genomic Structural Variation Detection in the Cloud with MapReduce}

\author[1,5]{Christopher W. Whelan \thanks{whelanch@ohsu.edu}}
\author[3,4]{Jeffrey Tyner}
\author[6]{Alberto L'Abbate}
\author[6]{Clelia Tiziana Storlazzi}
\author[4,5]{Lucia Carbone}
\author[1,2,5]{Kemal S\"onmez \thanks{sonmezk@ohsu.edu}}
\affil[1]{Institute on Development and Disability and Center for Spoken Language Understanding}
\affil[2]{Department of Medical Informatics \& Clinical Epidemiology}
\affil[3]{Program in Molecular and Cellular Biosciences}
\affil[4]{Behavioral Neuroscience Department}
\affil[5]{Oregon Health \& Science University, Portland, OR, USA}
\affil[6]{Department of Biology, University of Bari ``Aldo Moro'', Via G. Amendola 165/A, 70126, Bari, Italy}

\maketitle

 \begin{abstract}

The detection of genomic structural variations (SV) remains a difficult challenge in analyzing sequencing data, and the growing size and number of sequenced genomes have rendered SV detection a bona fide big data problem. MapReduce is a proven, scalable solution for distributed computing on huge data sets. We describe a conceptual framework for SV detection algorithms in MapReduce based on computing local genomic features, and use it to develop a deletion and insertion detection algorithm, Cloudbreak. On simulated and real data sets, Cloudbreak achieves accuracy improvements over popular SV detection algorithms, and genotypes variants from diploid samples. It provides dramatically shorter runtimes and the ability to scale to big data volumes on large compute clusters. Cloudbreak includes tools to set up and configure MapReduce (Hadoop) clusters on cloud services, enabling on-demand cluster computing. Our implementation and source code are available at \url{http://github.com/cwhelan/cloudbreak}. 

 \end{abstract}

\newpage

\section{Introduction}

Genomic structural variations (SVs) such as deletions, insertions, and inversions of DNA are widely prevalent in human populations and account for the majority of the bases that differ among normal human genomes \autocite{Mills:2011p1611, Conrad:2010ja}. However, detection of SVs with current high-throughput sequencing technology remains a difficult problem, with limited concordance between available algorithms and high false discovery rates \autocite{Mills:2011p1611}. Part of the problem stems from the fact that the signals indicating the presence of SVs are spread throughout large data sets, and integrating them to form an accurate detection measure is computationally difficult. As the volume of massively parallel sequencing data approaches ``big data'' scales, SV detection is becoming a time consuming component of genomics pipelines, and presents a significant challenge for research groups and clinical operations that may not be able to scale their computational infrastructure. Here we present a distributed software solution that is scalable and readily available on the cloud.

In other fields that have taken on the challenge of handling very large data sets, such as internet search, scalability has been addressed by computing frameworks that distribute processing to many compute nodes, each working on local copies of portions of the data. In particular, Google's MapReduce \autocite{Dean:2008p277} framework was designed to manage the storage and efficient processing of very large data sets across clusters of commodity servers. Hadoop is an open source project of the Apache Foundation which provides an implementation of MapReduce. MapReduce and Hadoop allow efficient processing of large data sets by executing tasks on nodes that are as close as possible the data they require, minimizing network traffic and I/O contention. The Hadoop framework has been shown to be effective in sequencing-related tasks including short read mapping \autocite{Schatz:2009p278}, SNP calling \autocite{Langmead:2009p1225}, and RNA-seq differential expression analysis \autocite{Langmead:2010p1268}.

Hadoop/MapReduce requires a specific programming model, however, which can make it difficult to design general-purpose algorithms for arbitrary sequencing problems like SV detection. MapReduce divides computation across a cluster into three phases. In the first phase, \emph{mappers} developed by the application programmer examine small blocks of data and emit a set of $\langle key, value \rangle$ pairs for each block examined. The framework then sorts the output of the mappers by key, and aggregates all values that are associated with each key. Finally, the framework executes \emph{reducers}, also created by the application developer, which process all of the values for a particular key and produce one or more outputs that summarize or aggregate those values.

Popular SV detection algorithms use three main signals present in high-throughput sequencing data sets \textcite{Alkan:2011p547}. Read-pair (RP) based methods use the distance between and orientation of the mappings of paired reads to identify the signatures of SVs \autocite{Campbell:2008p539,Chen:2009p3,Hormozdiari:2009p284,Sindi:2009gu,Korbel:2009dy}. Traditionally, this involves separating mappings into those that are \emph{concordant} or \emph{discordant}, where discordant mappings deviate from the expected insert size or orientation, and then clustering the discordant mappings to find SVs supported by multiple discordantly mapped read pairs. Read-depth (RD) approaches, in contrast, consider the changing depth of coverage of concordantly mapped reads along the genome \autocite{Abyzov:2011bk,Alkan:2009cr,Yoon:2009kb,Chiang:2009di}. Finally, split-read (SR) methods look for breakpoints by mapping portions of individual reads to different genomic locations \autocite{Wang:2011p1607,Ye:2009p2}.

Many RP methods consider only unambiguously discordantly mapped read pairs. Some approaches also include ambiguous mappings of discordant read pairs to improve sensitivity in repetitive regions of the genome \autocite{Hormozdiari:2009p284,Quinlan:2010gf}. Several recent RP approaches have also considered concordant read pairs, either to integrate RD signals for improved accuracy \autocite{Sindi:2012kk,Michaelson:2012fj,Chiara:2012ey}, or to eliminate the thresholds that separate concordant from discordant mappings and thus detect smaller events \autocite{Marschall:2012ek}. Increasing the number of read mappings considered, however, increases the computational burden of SV detection. 

Our goal is to leverage the strengths of the MapReduce computational framework in order to provide fast, accurate and readily scalable SV-detection pipelines. The main challenge in this endeavor is the need to separate logic into mappers and reducers, which makes it difficult to implement traditional RP-based SV detection approaches in MapReduce, particularly given the global clustering of paired end mappings at the heart of many RP approaches. MapReduce algorithms, by contrast, excel at conducting many independent calculations in parallel. In sequencing applications, for example, MapReduce based SNV-callers Crossbow \autocite{Langmead:2009p1225} and GATK \autocite{McKenna:2010p1051} perform independent calculations on partitions of the genome. SV approaches that are similarly based on local computations have been described: the RP-based SV callers MoDIL \autocite{Lee:2009da} and forestSV \autocite{Michaelson:2012fj} compute scores or features along the genome and then produce SV predictions from those features in a post-processing step. We will show that this strategy can be translated into the MapReduce architecture.

In this paper, we describe a framework for solving SV detection problems in Hadoop based on the computation of local genomic features from paired end mappings. In this framework we have developed a software package, Cloudbreak, that discovers genomic deletions up to 25,000bp long, and short insertions. Cloudbreak computes local features based on modeling the distribution of insert sizes at each genomic location as a Gaussian Mixture Model (GMM), an idea first implemented in MoDIL \autocite{Lee:2009da}. The use of Hadoop enables scalable implementations of this class of algorithm. We characterize our algorithm's performance on simulated and real data sets and compare its performance to those of several popular methods. 

Finally, and quite importantly from a practical point of view, our implementation of Cloudbreak provides the functionality to easily set up and configure Hadoop clusters on cloud service providers, making dynamically scalable distributed SV detection accessible to all whose computational needs demand it.

\section{Results}

\subsection{Cloudbreak: A Hadoop/MapReduce Software Package for SV Detection}

\begin{figure}
\centering
\includegraphics[width=.8\textwidth]{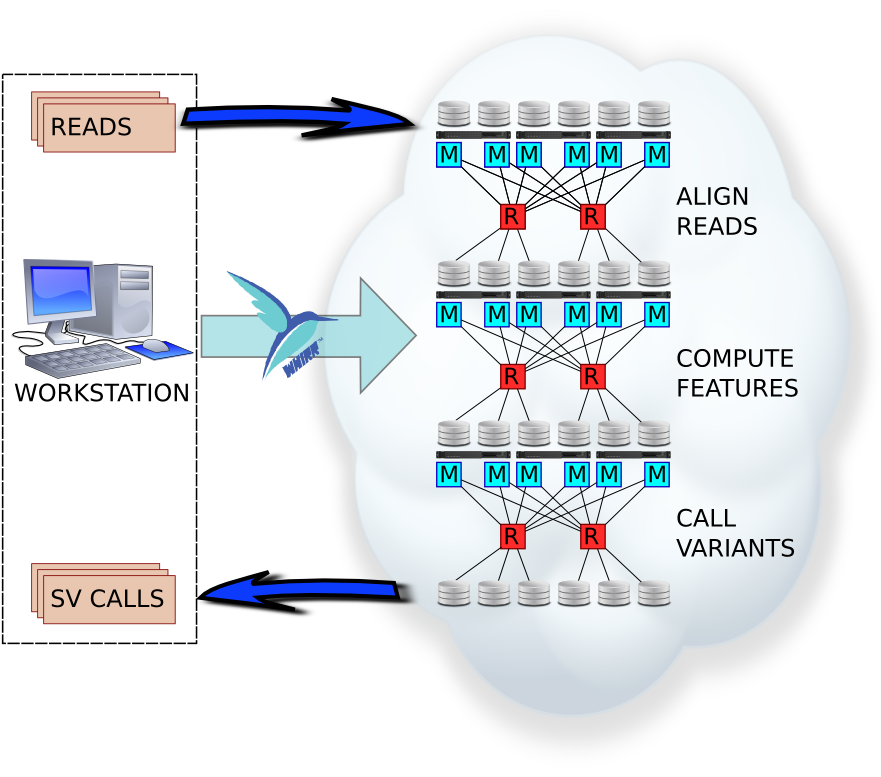}
\caption{An overview of the Cloudbreak workflow. Reads are first uploaded to a Hadoop cluster from local storage. Cloudbreak then executes three MapReduce jobs to process the data: 1) Mapping with sensitive settings. 2) Computation of features across the genome. 3) Calling structural variations based on the features computed in the previous step. Finally, SV predictions can be downloaded from the Hadoop cluster and examined. Cloudbreak can also use the Apache Whirr library to automatically provision Hadoop clusters on and deploy data to cloud providers such as Amazon Elastic Compute Cloud.}
\label{cloudbreak_workflow}
\end{figure}

Our framework for SV detection in MapReduce divides processing into three distinct MapReduce jobs (Figure \ref{cloudbreak_workflow}): a job that can align reads to the reference using a variety of mapping algorithms; a job that computes a set of features along the genome; and a job which calls structural variations based on those features. Our description of a general MapReduce SV detection algorithmic framework and how Cloudbreak is implemented within that framework are provided in Supplementary Algorithm \ref{cb_algo} and the Supplementary Materials; here we proceed with a high-level description of the Cloudbreak implementation.

Cloudbreak's alignment job can run a variety of alignment tools that report reads in SAM format (Supplementary Materials). In the map phase, mappers align reads in either single-end or paired-end mode to the reference genome in parallel, outputting mapping locations as values under a key identifying the read pair. In the reduce phase, the framework combines the reported mapping locations for the two ends of each read pair. This job can also be skipped in favor of importing a pre-existing set of mappings directly into the Hadoop cluster.

In the next job, Cloudbreak computes a set of features for each location in the genome. To begin, we tile the genome with small fixed-width, non-overlapping windows. For the experiments reported in this paper we use a window size of 25bp. Within each window, we examine the distribution of insert sizes of mappings that span that window, and compute features by fitting a GMM to that distribution (Supplementary Materials, Supplementary Figure \ref{insert_size_mixes}). To remove incorrect mappings, we use an adaptive quality cutoff for each genomic location and then perform an outlier-detection based noise reduction technique (Supplementary Materials); these procedures also allow us to process multiple mappings for each read if they are reported by the aligner.

Finally, the third MapReduce job is responsible for making SV calls based on these features. In this job, we search for contiguous blocks of genomic locations with similar features and merge them into individual insertion and deletion SV calls after applying noise reduction (Supplementary Materials).  Reducers process each chromosome in parallel after mappers find and organize its features. An illustration of the Cloudbreak algorithm working on a simple example is shown in Supplementary Figure \ref{algorithm_example}.

Cloudbreak can be executed on any Hadoop cluster; Hadoop abstracts away the details of cluster configuration, making distributed applications portable. In addition, our Cloudbreak implementation can leverage the Apache Whirr library to automatically create clusters with cloud service providers such as the Amazon Elastic Compute Cloud (EC2). This enables on demand provisioning of Hadoop clusters which can then be terminated when processing is complete, eliminating the need to invest in a standing cluster and allowing a model in which users can scale their computational infrastructure as their need for it varies over time.

\subsection{Tests with Simulated Data}

We compared the performance of Cloudbreak for detecting deletions and insertions to a selection of popular tools: the RP method BreakDancer \autocite{Chen:2009p3}, GASVPro, an RP method that integrates RD signals and ambiguous mappings \autocite{Sindi:2012kk}, the SR method Pindel \autocite{Ye:2009p2}, and the hybrid RP-SR method DELLY \autocite{Rausch:2012he}. DELLY produces two sets of calls, one based solely on RP signals, and the other based on RP calls that could be supported by SR evidence; we refer to these sets of calls as DELLY-RP and DELLY-SR. We also attempted to evaluate MoDIL on the same data. All of these methods detect deletions. Insertions can be detected by BreakDancer, Pindel, and MoDIL. See Methods for details on how reads were aligned and each program was invoked. For all alignments we used BWA \autocite{Li:2009p836}, although in testing Cloudbreak we have found that the choice of aligner, number of possible mapping locations reported, and whether the reads were aligned in paired-end or single-ended mode can have a variety of effects on the output of the algorithm (Supplementary Figure \ref{alignment_comparison}).

There is no available test set of real Illumina sequencing data from a sample that has a complete annotation of SVs. Therefore, testing with simulated data is important to fully characterize an algorithm's performance characteristics. On the other hand, any simulated data should contain realistic SVs that follow patterns observed in real data. Therefore, we took one of the most complete lists of SVs from an individual, the list of homozygous insertions and deletions from the genome of J. Craig Venter \autocite{Levy:2007fb}, and used it to simulate a 30X read coverage data set for a diploid human Chromosome 2 with a mix of homozygous and heterozygous variants, with 100bp reads and a mean fragment size of 300bp (See Methods).

\begin{figure}
\centering
\includegraphics[width=1\textwidth]{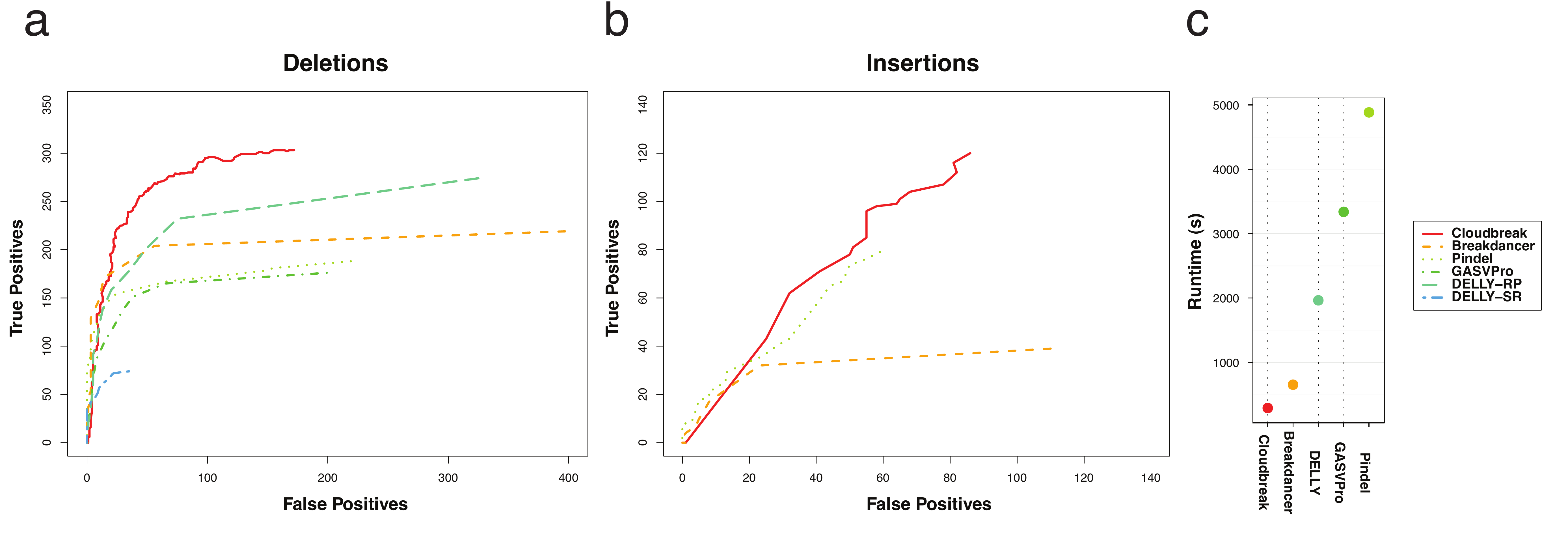}
\caption{Accuracy and runtime performance on a simulated data set. (a) Receiver Operating Characteristic (ROC) curves showing the specificity and sensitivity of each tool to deletions larger than 40bp on a simulated set of reads giving diploid coverage of 30X on human chromosome 2. Deletions from the Venter genome were randomly added to one or both haplotypes. Each point on a curve represents a different threshold on the confidence of predictions made by that tool. Thresholds vary by: Cloudbreak - likelihood ratio; BreakDancer, DELLY, GASVPro - number of supporting read pairs; Pindel - simple score. (b) ROC curves for insertion predictions. (c) Runtimes for each tool, not including alignment time, parallelized when possible (see Supplementary Material).}
\label{chr2CombinedRoc}
\end{figure}

Figure \ref{chr2CombinedRoc} shows Receiver Operating Characteristic (ROC) curves of the performance of each algorithm at identifying variants larger than 40bp on the simulated data set, as well as the runtimes of the approaches tested, excluding alignment. See Methods for a description of how we identified correct predictions. All approaches show excellent specificity at high thresholds in this simulation. Cloudbreak provides the greatest specificity for deletions at higher levels of sensitivity, followed by DELLY. For insertions, Cloudbreak's performance is similar to or slightly better than Pindel. Cloudbreak's runtime is half that of BreakDancer, the next fastest tool, processing the simulated data in under six minutes. (Of course, Cloudbreak uses many more CPUs as distributed algorithm. See Supplementary Material and Supplementary Table \ref{runtimes} for a discussion of runtimes and parallelization.) The output which we obtained from MoDIL did not have a threshold that could be varied to correlate with the trade-off between precision and recall and therefore it is not included in ROC curves; in addition, MoDIL ran for 52,547 seconds using 250 CPUs in our cluster. Apart from the alignment phase, which is embarrassingly parallel, the feature generation job is the most computationally intensive part of the Cloudbreak workflow. To test its scalability we measured its runtime on Hadoop clusters made up of varying numbers of nodes and observed that linear speedups can be achieved in this portion of the algorithm by adding additional nodes to the cluster until a point of diminishing returns is reached (Supplementary Figure \ref{scalability}).

Choosing the correct threshold to set on the output of an SV calling algorithm can be difficult. The use of simulated data and ROC curves allows for some investigation of the performance characteristics of algorithms at varying thresholds. First, we characterized the predictions made by each algorithm at the threshold that gives them maximum sensitivity. For Cloudbreak we chose an operating point at which marginal improvements in sensitivity became very low. The results are summarized in Table \ref{chr2DeletionAndInsertionPredsMaxSensitivity}. MoDIL and Cloudbreak exhibited the greatest recall for deletions. Cloudbreak has also has high precision at this threshold, and discovers many small variants. For insertions, Cloudbreak has the highest recall, although recall is low for all four approaches. Cloudbreak again identifies many small variants. Pindel is the only tool which can consistently identify large insertions, as insertions larger than the library insert size do not produce mapping signatures detectable by RP mapping. We also used the ROC curves to characterize algorithm performance when a low false discovery rate is required. Supplementary Table \ref{chr2DeletionPredsFDR10} shows the total number of deletions found by each tool when choosing a threshold that gives an FDR closest to 10\% based on the ROC curve. At this more stringent cutoff, Cloudbreak identifies more deletions in every size category than any other tool. Insertions performance never reached an FDR of 10\% for any threshold, so insertion predictions are not included in this table. We also examined Cloudbreak's ability to detect events in repetitive regions of the genome, and found that it was similar to the other methods tested (Supplementary Tables \ref{deletionRepmaskpreds} and \ref{insertionRepmaskpreds}).

It should be noted that the methods tested here vary in their breakpoint resolution (Supplementary Figure \ref{breakpoint_resolution}): SR methods have higher resolution than RP methods. Cloudbreak sacrifices additional resolution by dividing the genome into 25bp windows; we believe, however, that increasing sensitivity and specificity is of greatest utility, especially given the emergence of pipelines in which RP calls are validated \emph{in silico} by local assembly.

\begin{table}[t]
\begin{center}
\resizebox{\textwidth}{!}{
\begin{tabular}{r|rrr|rrrrr}
  \cline{2-9}
   &                     & Prec. & Recall & 40-100bp  & 101-250bp  & 251-500bp & 501-1000bp & $>$ 1000bp \\ 
\hline
\multirow{7}{*}{\begin{sideways}Deletions\end{sideways}} & Total Number &          &           & 224 &  84 & 82 &  31 & 26\\ 
  \hline
\cline{2-9}
&  Cloudbreak    &  0.638 & \textbf{0.678} & \textbf{153} (9)  & 61 (0) &  62 (0) & 12 (0) & 15 (0) \\ 
&  BreakDancer   &  0.356 & 0.49 & 89 (0)  & 54 (0) &  53 (0) & 8 (0) & 15 (0) \\ 
&  GASVPro        & 0.146 & 0.432 & 83 (2)  & 32 (0) &  55 (0) & 8 (0) & 15 (0) \\ 
&  DELLY-RP           & 0.457 & 0.613 & 114 (3)  & \textbf{68} (0) &  \textbf{66} (0) & 9 (1) & 17 (0) \\ 
&  DELLY-SR           & \textbf{0.679} & 0.166 & 0 (0)  & 3 (0) &  49 (0) & 6 (0) & 16 (0) \\ 
&  Pindel           & 0.462 & 0.421 & 96 (\textbf{11})  & 24 (0) &  48 (0) & 5 (0) & 15 (0)\\ 
&  MoDIL           & 0.132  & 0.66 & 123 (6)  & 66 (\textbf{3}) &  \textbf{66} (\textbf{11}) & \textbf{17} (\textbf{7}) & \textbf{23} (\textbf{8})\\ 
   \hline
\multirow{5}{*}{\begin{sideways}Insertions\end{sideways}} & Total Number &          &           & 199 &  83 & 79 &  21 & 21\\ 
\cline{2-9}
&  Cloudbreak   &0.451 & \textbf{0.305}  & \textbf{79} (\textbf{32})  & \textbf{32} (\textbf{18}) &  \textbf{11} (8) & 1 (0) & 0 (0) \\ 
&  BreakDancer & 0.262 & 0.0968  & 23 (5)  & 14 (5) &  2 (1) & 0 (0) & 0 (0) \\ 
&  Pindel          & \textbf{0.572} & 0.196 & 52 (25)  & 5 (1) &  10 (\textbf{9}) & \textbf{3} (\textbf{2}) & \textbf{9} (\textbf{9})\\ 
&  MoDIL          & 0.186 & 0.0521 & 14 (1)  & 4 (0) &  1 (0) & 2 (\textbf{2}) & 0 (0)\\ 
\hline
\end{tabular}
}
\end{center}
\caption{The number of simulated deletions and insertions in the 30X diploid chromosome 2 with Venter indels found by each tool at maximum sensitivity, as well as the number of those variants that were discovered exclusively by each tool (in parentheses). The total number of variants in each size class in the true set of deletions and insertions is shown in the first row of each section.}
\label{chr2DeletionAndInsertionPredsMaxSensitivity}
\end{table}

\subsection{Tests with Biological Data}

We downloaded a set of reads from Yoruban individual NA18507, experiment ERX009609, from the Sequence Read Archive. This sample was sequenced by Illumina Inc. on the Genome Analyzer II platform with 100bp paired end reads and a mean fragment size (minus adapters) of 300bp, with a standard deviation of 15bp, to a depth of approximately 37X coverage. To create a gold standard set of insertions and deletions to test against, we pooled annotated variants discovered by three previous studies on the same individual. These included data from the Human Genome Structural Variation Project reported by \textcite{Kidd:2008p926}, a survey of small indels conducted by \textcite{Mills:2011fi}, and insertions and deletions from the merged call set of the phase 1 release of the 1000 Genomes Project \autocite{GenomesProjectConsortium:2012co} which were genotyped as present in NA18507. We merged any overlapping calls of the same type into the region spanned by their unions. We were unable to run MoDIL on the whole-genome data set due to the estimated runtime and storage requirements.

\begin{figure}
\centering
\includegraphics[width=1\textwidth]{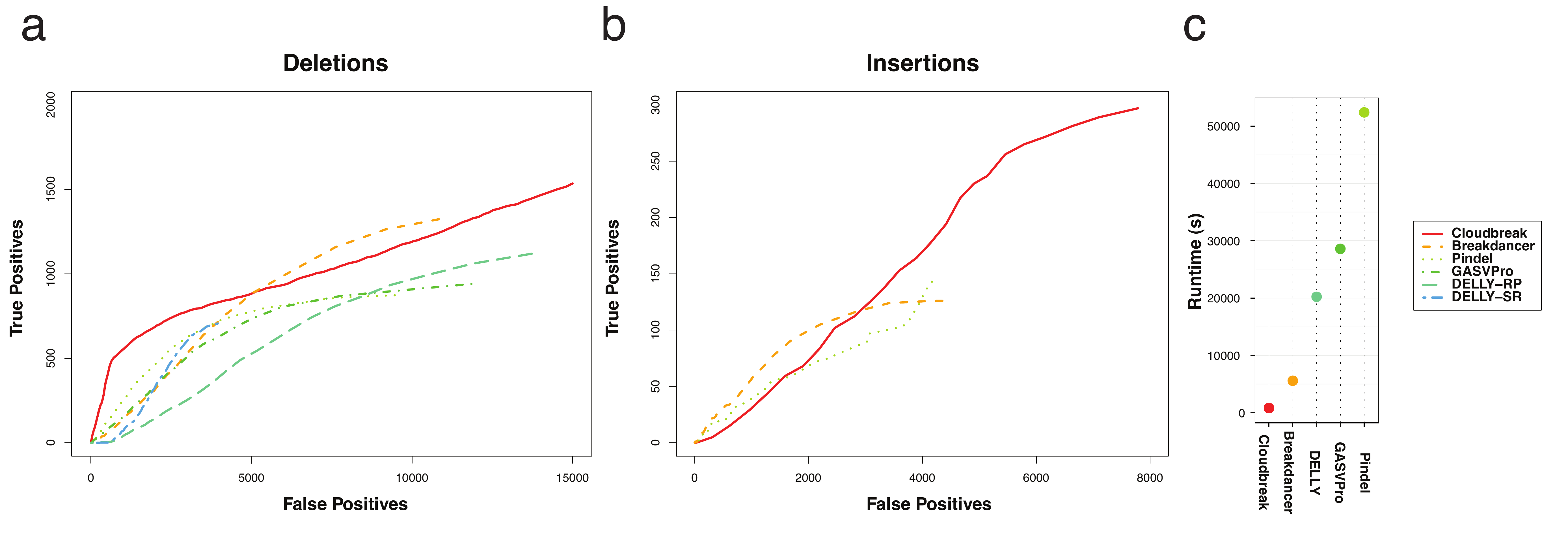}
\caption{Accuracy and performance on the 37X NA18507 sample. (a) ROC curve for deletion prediction performance, tested against the combined gold standard sets of deletions taken from \textcite{Kidd:2008p926}, \textcite{Mills:2011fi}, and \textcite{GenomesProjectConsortium:2012co}. (b) ROC curve for insertion prediction performance. (c) Runtimes for each tool, not including alignment time, parallelized when possible (see Supplementary Material). }
\label{NA18507CombinedRoc}
\end{figure}

Figure \ref{NA18507CombinedRoc} shows the performance of each algorithm at detecting events larger than 40bp on the NA18507 data set. All algorithms show far less specificity for the gold standard set than they did in the single chromosome simulation, although it is difficult to tell how much of the difference is due to the added complexity of real data and a whole genome, and how much is due to missing variants in the gold standard set that are actually present in the sample. For deletions, Cloudbreak is the best performer at the most stringent thresholds, and has the highest or second highest precision at higher sensitivity levels. Cloudbreak has comparable accuracy for insertions to other tools, and can identify the most variants at higher levels of sensitivity. Cloudbreak processes the sample in under 15 minutes on our cluster, more than six times as fast as the next fastest program, BreakDancer, even when BreakDancer is run in parallel for each chromosome on different nodes in the cluster (see Methods and Supplementary Material).

Given the high number of novel predictions made by all tools at maximum sensitivity, we decided to characterize performance at more stringent thresholds. We examined the deletion predictions made by each algorithm using the same cutoffs that yielded a 10\% FDR on the simulated chromosome 2 data set, adjusted proportionally for the difference in coverage from 30X to 37X. For insertions, we used the maximum sensitivity thresholds for each tool due to the high observed FDRs in the simulated data. Precision and recall at these thresholds, as well as the performance of each algorithm at predicting variants of each size class, is shown in Table \ref{NA18507DeletionAndInsertionPreds}. For deletions, Cloudbreak has the greatest sensitivity of any tool, identifying the most variants in each size class. Pindel exhibits the highest precision with respect to the gold standard set. For insertions, Pindel again has the highest precision at maximum sensitivity, while Cloudbreak has by far the highest recall.

\begin{table}
\begin{center}
\resizebox{\textwidth}{!}{
\begin{tabular}{r|rrr|rrrrr}
  \cline{2-9}
&  & Prec. & Recall & 40-100bp & 101-250bp & 251-500bp & 501-1000bp & $>$ 1000bp \\ 
\hline
\multirow{6}{*}{\begin{sideways}Deletions\end{sideways}} & Total Number & & & 7,462 & 240 & 232 & 147 & 540 \\
  \hline
\cline{2-9}
& Cloudbreak & 0.0943 & \textbf{0.17} & \textbf{573} (\textbf{277})  & \textbf{176} (\textbf{30}) &  \textbf{197} (\textbf{18}) & \textbf{121} (\textbf{6}) & \textbf{399} (\textbf{24}) \\ 
& BreakDancer & 0.137 & 0.123 & 261 (29)  & 136 (3) &  178 (0) & 114 (0) & 371 (0) \\  
&  GASVPro & 0.147 & 0.0474 & 120 (21)  & 40 (2) &  85 (0) & 36 (0) & 128 (0) \\ 
&  DELLY-RP & 0.0931 & 0.1 & 143 (6)  & 128 (3) &  167 (1) & 103 (0) & 323 (1) \\ 
&  DELLY-SR & 0.153 & 0.0485 & 0 (0)  & 26 (0) &  123 (0) & 66 (0) & 203 (0) \\ 
&  Pindel & \textbf{0.179} & 0.0748 & 149 (8)  & 61 (0) &  149 (0) & 69 (1) & 217 (0) \\ 
\hline
\multirow{4}{*}{\begin{sideways}Insertions\end{sideways}} & Total Number & & & 536 & 114 & 45 & 1 & 0 \\
\cline{2-9}
& Cloudbreak & 0.0323 & \textbf{0.455} & \textbf{265} (\textbf{104})  & \textbf{49} (\textbf{24}) &  3 (1) & 0 (0)  & 0 (0)  \\ 
& BreakDancer & 0.0281 & 0.181 & 97 (10)  & 27 (5) &  2 (1) & 0 (0) & 0 (0) \\  
&  Pindel & \textbf{0.0387} & 0.239 & 144 (45)  & 14 (7) &  \textbf{7} (\textbf{6}) & \textbf{1} (\textbf{1}) &  0 (0) \\ 
\hline
\end{tabular}
}
\end{center}
\caption{The precision and recall with respect to the gold standard set of deletions and insertions for each tool on the NA18507 data, as well as the number of variants found in each size class found. Exclusive predictions are in parentheses. For deletions, the same cutoffs were used as for the simulated data as in Supplementary Table \ref{chr2DeletionPredsFDR10}, adjusted for the difference in coverage from 30X to 37X. For insertions, the maximum sensitivity cutoff was used.}
\label{NA18507DeletionAndInsertionPreds}
\end{table}

\subsection{Performance on a Low-Coverage Cancer Data Set}

We also tested Cloudbreak on a sequencing data set obtained from a patient with acute myeloid leukemia (AML). This data set consisted of 76bp paired end reads with a mean insert size of 285bp and standard deviation of 50bp, yielding sequence coverage of 5X and physical coverage of 8X. Using a pipeline consisting of Novoalign, BreakDancer, and a set of custom scripts for filtering and annotating candidate SVs, we had previously identified a set of variants present in this sample and validated several using PCR, including 8 deletions. Cloudbreak was able to identify all 8 of the validated deletions, showing that it is still sensitive to variants even when using lower coverage data sets with a greater variance of insert sizes. The variants identified include deletions in the gene CTDSPL/RBPS3, an AML tumor suppressor \autocite{Zheng:2012kk}, and NBEAL1, a gene up-regulated in some cancers \autocite{Chen:2004jo}. We are currently investigating these deletions to determine their functional impact on this patient. 

\subsection{Genotyping Variants}

Because Cloudbreak explicitly models zygosity in its feature generation algorithm, it can predict the genotypes of identified variants. We tested this on both the simulated and NA18507 data sets. For the NA18507 data set, we considered the deletions from the 1000 Genomes Project, which had been genotyped using the population-scale SV detection algorithm Genome STRiP \autocite{Handsaker:2011ki}. Cloudbreak was able to achieve 92.7\% and 95.9\% accuracy in predicting the genotype of the deletions it detected at our 10\% FDR threshold in the simulated and real data sets, respectively. Supplementary Table \ref{deletionGenotypeaccuracy} shows confusion matrices for the two samples using this classifier. None of the three input sets that made up the gold standard for NA18507 contained a sufficient number of insertions that met our size threshold and also had genotyping information. Of the 123 insertions detected by Cloudbreak on the simulated data set, 43 were heterozygous. Cloudbreak correctly classified 78 of the 80 homozygous insertions and 31 of the 43 heterozygous insertions, for an overall accuracy of 88.6\%.

\section{Discussion}\label{Discussion}

Over the next few years, due to advances in sequencing technology, genomics data are expected to increase in volume by several orders of magnitude, expanding into the realm referred to as ``big data''. In addition, the usage of existing information will increase drastically as research in genomics grows and translational applications are developed, and the data sets are reprocessed and integrated into new pipelines. In order to capitalize on this emerging wealth of genome data, novel computational solutions that are capable of scaling with the increasing number and size of these data sets will have to be developed. 

Among big data infrastructures, MapReduce is emerging as a standard framework for distributing computation across compute clusters. In this paper, we introduced a novel conceptual framework for SV detection algorithms in MapReduce, based on computing local genomic features. This framework provides a scalable basis for developing SV detection algorithms, as demonstrated by our development of an algorithm for detecting insertions and deletions based on fitting a GMM to the distribution of mapped insert sizes spanning each genomic location.

On simulated and real data sets, our approach exhibits high accuracy when compared to popular SV detection algorithms that run on traditional clusters and servers. Detection of insertions and deletions is an important area of research; \textcite{Mills:2011fi} recently identified over 220 coding deletions in a survey of a large number of individuals, and they note that such variants are likely to cause phenotypic variation in humans. More SV annotations are also required to assess the impact of SVs on non-coding elements \autocite{Mu:2011br}.

In addition to delivering state-of-the-art performance in a RP-based SV detection tool, our approach offers a basis for developing a variety of SV algorithms that are capable of running in a MapReduce pipeline with the power to process vast amounts of data in a cloud or commodity server setting. With the advent of cloud service providers such as Amazon EC2, it is becoming easy to instantiate on-demand Hadoop compute clusters. Having computational approaches that can harness this capability will become increasingly important for researchers or clinicians who need to analyze increasing amounts of sequencing data.

\newpage

\section{Methods}\label{discussion}

\subsection{Cloudbreak Implementation}
Cloudbreak is a native Java Hadoop application. We deployed Cloudbreak on a 56-node cluster running the Cloudera CDH3 Hadoop distribution, version 0.20.2-cdh3u4. We use snappy compression for MapReduce data. Hadoop's distributed cache mechanism shares the executable files and indices needed for mapping tasks to the nodes in the cluster. To execute the other tools in parallel mode we wrote simple scripts to submit jobs to the cluster using the HTCondor scheduling engine (\url{http://research.cs.wisc.edu/htcondor/}) with directed acyclic graphs to describe dependencies between jobs. 

\subsection{Cloudbreak Genotyping}
We used the parameters of the fit GMM to infer the genotype of each predicted variant. Assuming that our pipeline is capturing all relevant read mappings near the locus of the variant, the genotype should be indicated by the estimated parameter $\alpha$, the mixing parameter that controls the weight of the two components in the GMM. We setting a simple cutoff of .35 on the average value of $\alpha$ for each prediction to call the predicted variant homozygous or heterozygous. We used the same cutoff for deletion and insertion predictions.

\subsection{Read Simulation}
Since there are relatively few heterozygous insertions and deletions annotated in the Venter genome, we used the set of homozygous indels contained in the HuRef data (\texttt{HuRef.homozygous\_indels.061109.gff}) and randomly assigned each variant to be either homozygous or heterozygous. Based on this genotype, we applied each variant to one or both of two copies of the human GRCh36 chromosome 2 reference sequence. We then simulated paired Illumina reads from these modified references using \emph{dwgsim} from the DNAA software package (\url{http://sourceforge.net/apps/mediawiki/dnaa/}). We simulated 100bp reads with a mean fragment size of 300bp and a standard deviation of 30bp, and generated 15X coverage for each modified sequence. Pooling the reads from both simulations gives 30X coverage for a diploid sample with a mix of homozygous and heterozygous insertions and deletions.

\subsection{Read Alignments}
Simulated reads were aligned to hg18 chromosome 2, and NA18507 reads were aligned to the hg19 assembly. Alignments for all programs, unless otherwise noted, were aligned using BWA \autocite{Li:2009p836} \texttt{aln} version 0.6.2-r126, with parameter \texttt{-e 5} to allow for longer gaps in alignments due to the number of small indels near the ends of larger indels in the Venter data set. GASVPro also accepts ambiguous mappings; we extracted read pairs that did not align concordantly with BWA and re-aligned them with Novoalign V2.08.01, with parameters \texttt{-a -r -Ex 1100 -t 250}. 

\subsection{SV Tool Execution}
We ran BreakDancer version 1.1\_2011\_02\_21 in single threaded mode by first executing \texttt{bam2cfg.pl} and then running \texttt{breakdancer\_max} with the default parameter values.  To run Breakdancer in parallel mode we first ran \texttt{bam2cfg.pl} and then launched parallel instances of \texttt{breakdancer\_max} for each chromosome using the \texttt{-o} parameter. We ran DELLY version 0.0.9 with the \texttt{-p} parameter and default values for other parameters. For the parallel run of DELLY we first split the original BAM file with BamTools \autocite{Barnett:2011hm}, and then ran instances of DELLY in parallel for each BAM file. We ran GASVPro version 1.2 using the \texttt{GASVPro.sh} script and default parameters. Pindel 0.2.4t was executed with default parameters in single CPU mode, and executed in parallel mode for each chromosome using the \texttt{-c} option. We executed MoDIL with default parameters except for a \texttt{MAX\_DEL\_SIZE} of 25000, and processed it in parallel on our cluster with a step size of 121475.

\subsection{SV Prediction Evaluation}
We use the following criteria to define a true prediction given a gold standard set of deletion and insertion variants to test against: A predicted deletion is counted as a true positive if a) it overlaps with a deletion from the gold standard set, b) the length of the predicted call is within 300bp (the library fragment size in both our real and simulated libraries) of the length of the true deletion, and c) the true deletion has not been already been discovered by another prediction from the same method. For evaluating insertions, each algorithm produces insertion predictions that define an interval in which the insertion is predicted to have occurred with start and end coordinates $s$ and $e$ as well as the predicted length of the insertion, $l$. The true insertions are defined in terms of their actual insertion coordinate $i$ and their actual length $l_a$. Given this information, we modify the overlap criteria in a) to include overlaps of the intervals $\langle s,\max{\left(e,s+l\right)} \rangle$ and $\langle i,i+l_a \rangle$. In this study we are interested in detecting events larger than 40bp, because with longer reads, smaller events can be more easily discovered by examining gaps in individual reads. Both Pindel and MoDIL make many calls with a predicted event size of under 40bp, so we remove those calls from the output sets of those programs. Finally, we exclude from consideration calls from all approaches that match a true deletion of less than 40bp where the predicted variant length is less than or equal to 75bp in length.

\subsection{Preparation of AML Sample}

Peripheral blood was collected from a patient with acute myelomonocytic leukemia (previously designated acute myeloid leukemia FAB M4) under a written and oral informed consent process reviewed and approved by the Institutional Review Board of Oregon Health \& Science University. Known cytogenetic abnormalities associated with this specimen included trisomy 8 and internal tandem duplications within the FLT3 gene. Mononuclear cells were separated on a Ficoll gradient, followed by red cell lysis. Mononuclear cells were immunostained using antibodies specific for CD3, CD14, CD34, and CD117 (all from BD Biosciences) and cell fractions were sorted using a BD FACSAria flow cytometer. Cell fractions isolated included T-cells (CD3-positive), malignant monocytes (CD14-positive), and malignant blasts (CD34, CD117-positive). We sequenced CD14+ cells on an Illumina Genome Analyzer II, producing 128,819,200 paired-end reads.

\subsection{Validation of AML Deletions by PCR}

Deletions identified in the AML dataset were validated by PCR. SVs to validate were selected based on calls made by BreakDancer, relying on the score and the number of the reads supporting each single event. Appropriate primers were designed with an internet-based interface, Primer3 (\url{http://frodo.wi.mit.edu/}), considering the chromosome localization and orientation of the interval involved in the candidate rearrangement. The primers were checked for specificity using the BLAT tool of the UCSC Human Genome Browser (\url{http://genome.ucsc.edu/cgi-bin/hgBlat}). All the primer pairs were preliminarily tested on the patient genomic DNA and a normal genomic DNA as control. The PCR conditions were as follows: 2 min at 95\degree C followed by 35 cycles of 30 sec at 95\degree C, 20 sec at 60\degree C, and 2 min at 72\degree C. All the obtained PCR products were sequenced and analyzed by BLAT for sequence specificity.

\newpage

\section{Author Contributions}

C.W.W. and K.S. designed the algorithmic approaches. C.W.W. implemented the software and conducted the evaluations. J.T. obtained the AML sample and sorted the cells. L.C. sequenced the AML sample. C.T.S and A.L.A. performed PCR validations on the AML sample. C.W.W. and K.S. wrote the manuscript.

\section*{Acknowledgments}

We would like to thank Izhak Shafran at the Center for Spoken Language Understanding for advice, support, and shared computational resources, and Bob Handsaker for helpful conversations. We also thank reviewers for the Third Annual RECOMB Satellite Workshop on Massively Parallel Sequencing (RECOMB-seq) for their comments on a preliminary description of Cloudbreak. C.T.S. and A.L.A. are supported by the Italian Association for Cancer Research (AIRC). Sequencing of the AML sample was supported by the Oregon Medical Research Foundation. 

\printbibliography

\setcounter{table}{0}
\setcounter{figure}{0}
\renewcommand{\figurename}{Supplementary Figure}
\renewcommand{\tablename}{Supplementary Table}

\newfloat{suppalgorithm}{p}{cap}
\floatname{suppalgorithm}{Supplementary Algorithm}

\newrefsection

\begin{figure}
\centering
\includegraphics[width=1\textwidth]{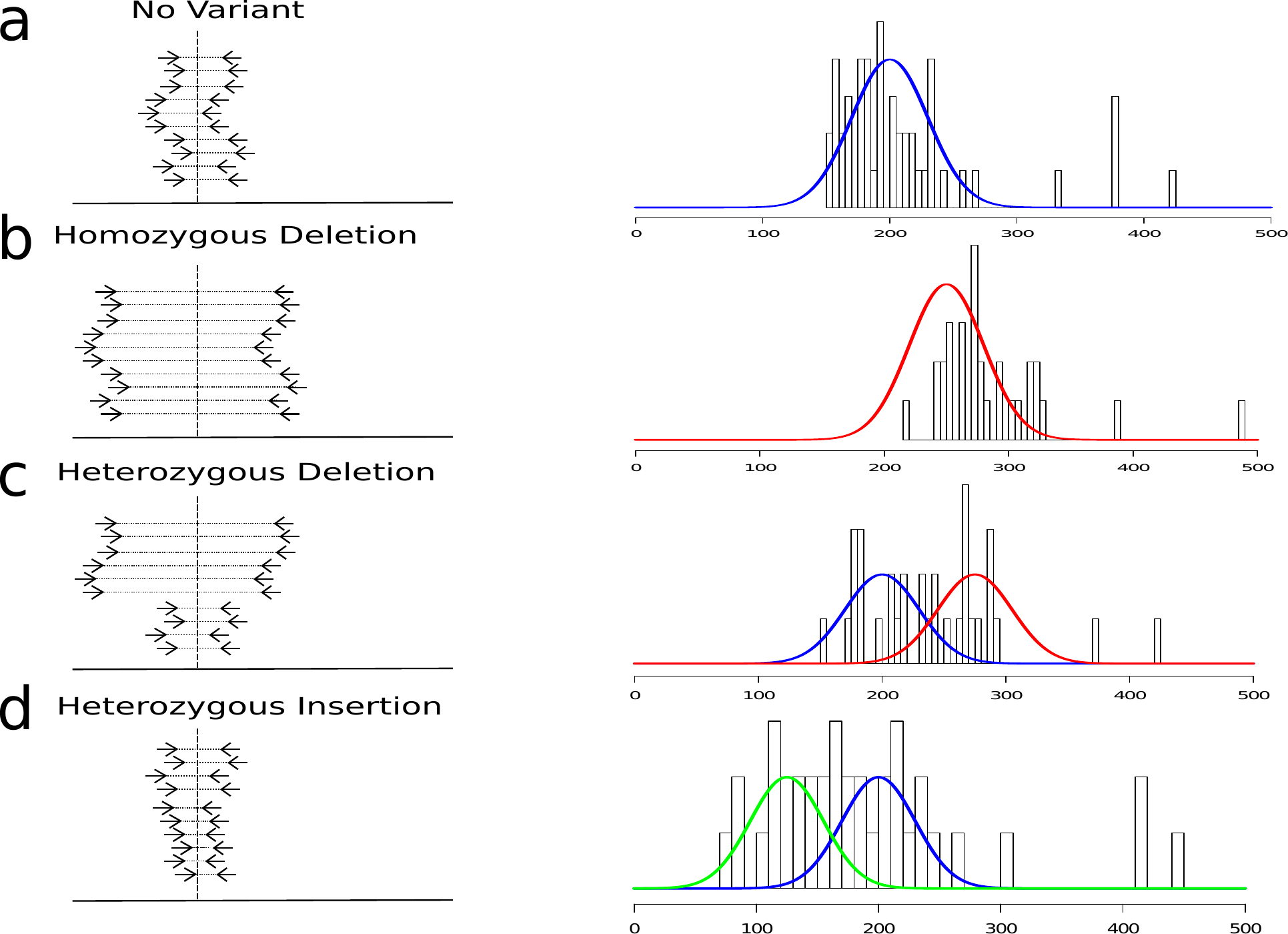}
\caption{Illustration of insert size mixtures at individual genomic locations. A) There is no variant present at the location indicated by the vertical line (left), so the mix of insert sizes (right) follows the expected distribution of the library centered at 200bp, with a small amount of noise coming from low-quality mappings. B) A homozygous deletion of 50bp at the location has shifted the distribution of observed insert sizes. C) A heterozygous deletion at the location causes a mixture of normal and long insert sizes to be detected. D) A heterozygous small insertion shifts a portion of the mixture to have lower insert sizes.}
\label{insert_size_mixes}
\end{figure}

\clearpage

\begin{figure}[h]
\centering
\includegraphics[width=1\textwidth]{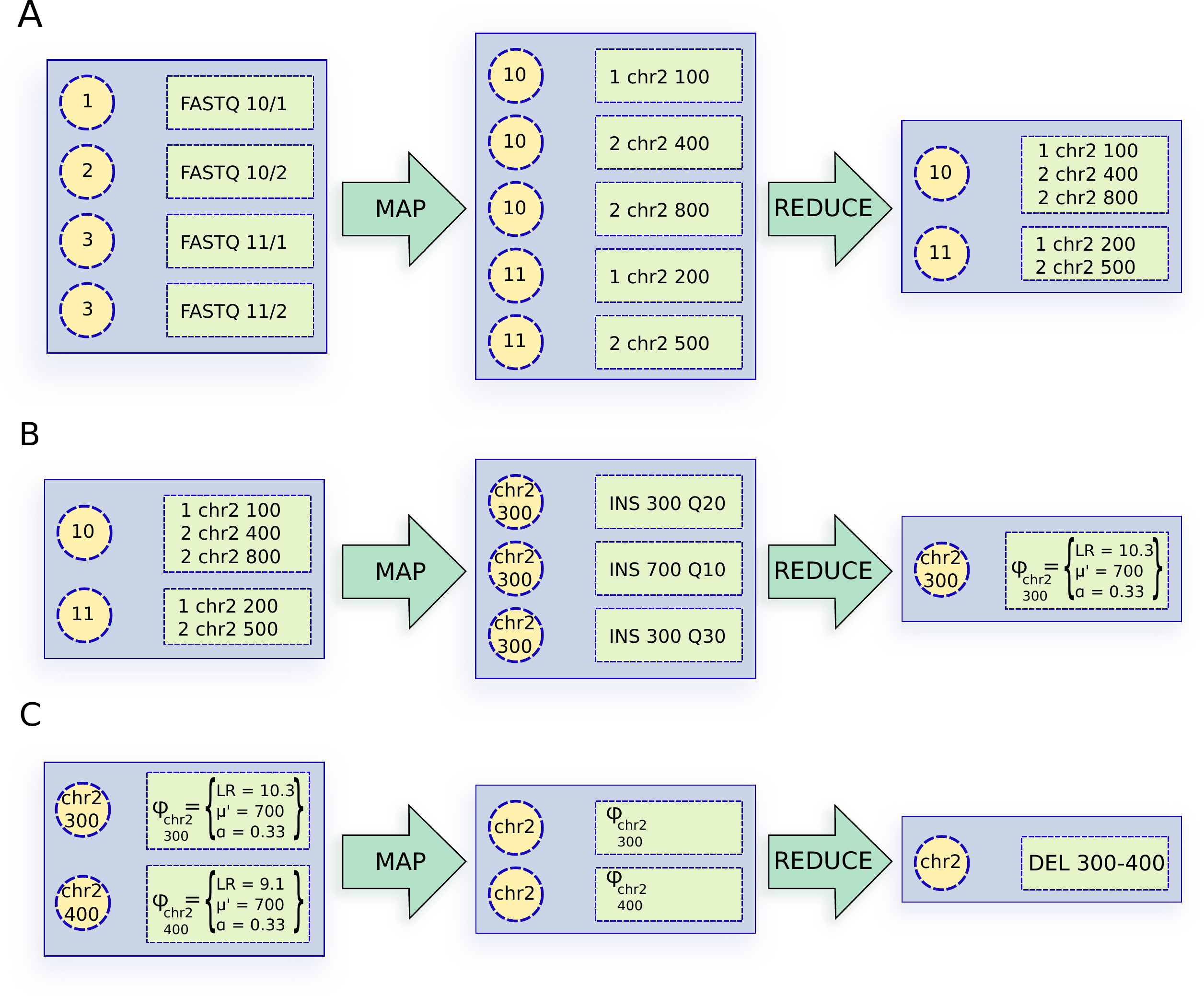}
\caption{An example of the Cloudbreak MapReduce algorithm. A) In the first MapReduce job, mappers scan input reads in FASTQ format and execute an alignment program in either paired-end or single-ended mode to generate read mappings. Reducers gather all alignments for both reads in each pair. B) In the second MapReduce job, mappers first emit information about each read pair (in this case the insert size and quality) under keys indicating the genomic location spanned by that pair. Only one genomic location is diagrammed here for simplicity. Reducers then compute features for each location on the genome by fitting a GMM to the distribution of spanning insert sizes. C) Mappers group all emitted features by their chromosome, and reducers find contiguous blocks of features that indicate the presence of a deletion.}
\label{algorithm_example}
\end{figure}

\clearpage

\begin{figure}
\centering
\includegraphics[width=1\textwidth]{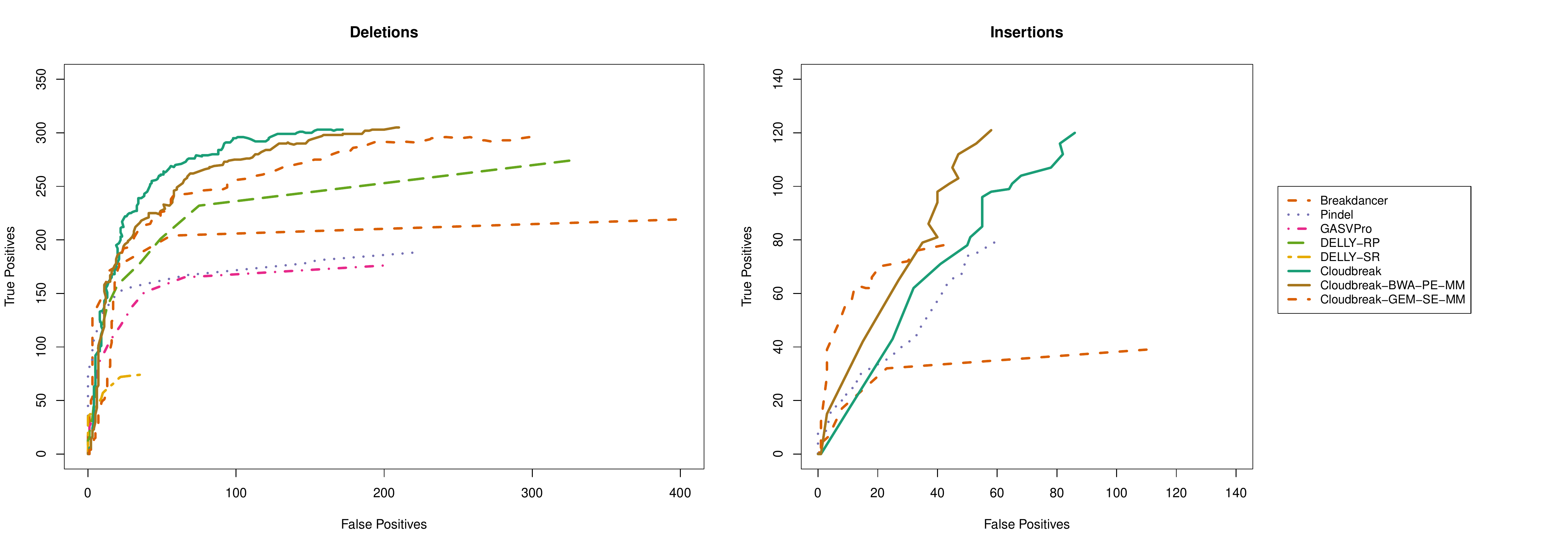}
\caption{Cloudbreak performance on the chromosome 2 simulation using different alignment strategies. ROC curves show the number of true positives and false positives for each operating point for deletions and insertions. The Cloudbreak alignment strategies are: 1) ``Cloudbreak'': Alignments generated with BWA in paired-end mode, reporting the best hit for each pair. 2) ``Cloudbreak-BWA-PE-MM'': Alignments generated with BWA in paired-end mode, reporting up to 25 additional hits for each mapping in SAM format using the \texttt{-n} and \texttt{-N} parameters for \texttt{bwa sampe} and the script \texttt{xa2multi.pl}. 3) ``Cloudbreak-GEM-SE-MM'': Alignments generated by running the GEM aligner in single-ended mode, reporting up to 1000 additional hits per alignment. GEM was executed in parallel using Hadoop tasks which wrap GEM version 1.362 (beta), with parameters \texttt{-e 6 -m 6 -s 2 -q ignore -d 1000 --max-big-indel-length 0},  requesting all hits for a read that are within an edit distance of 6 of the reference, within 2 strata of the best hit, with a maximum of 1000 possible alignments reported for each read. Considering multiple mappings improves Cloudbreak's specificity for insertions but decreases sensitivity to deletions.}
\label{alignment_comparison}
\end{figure}

\clearpage

\begin{figure}
\centering
\includegraphics[width=1\textwidth]{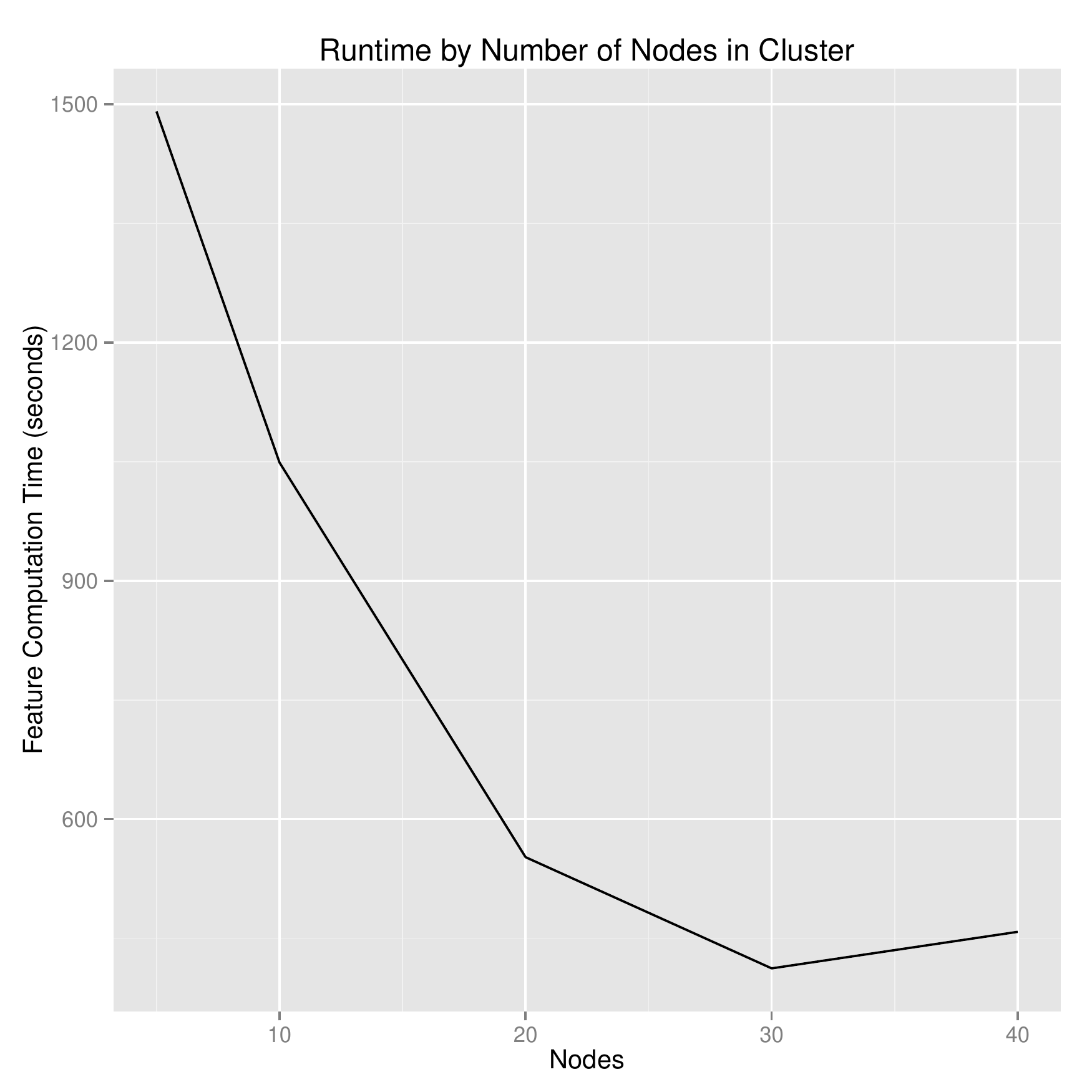}
\caption{Scalability of the Cloudbreak algorithm. Runtime of the Cloudbreak feature generation job for the simulated Chromosome 2 data is shown on Hadoop clusters consisting of varying numbers of compute nodes. Clusters were created in the Amazon Elastic Compute Cloud.}
\label{scalability}
\end{figure}

\clearpage

\begin{figure}
\centering
\includegraphics[width=1\textwidth]{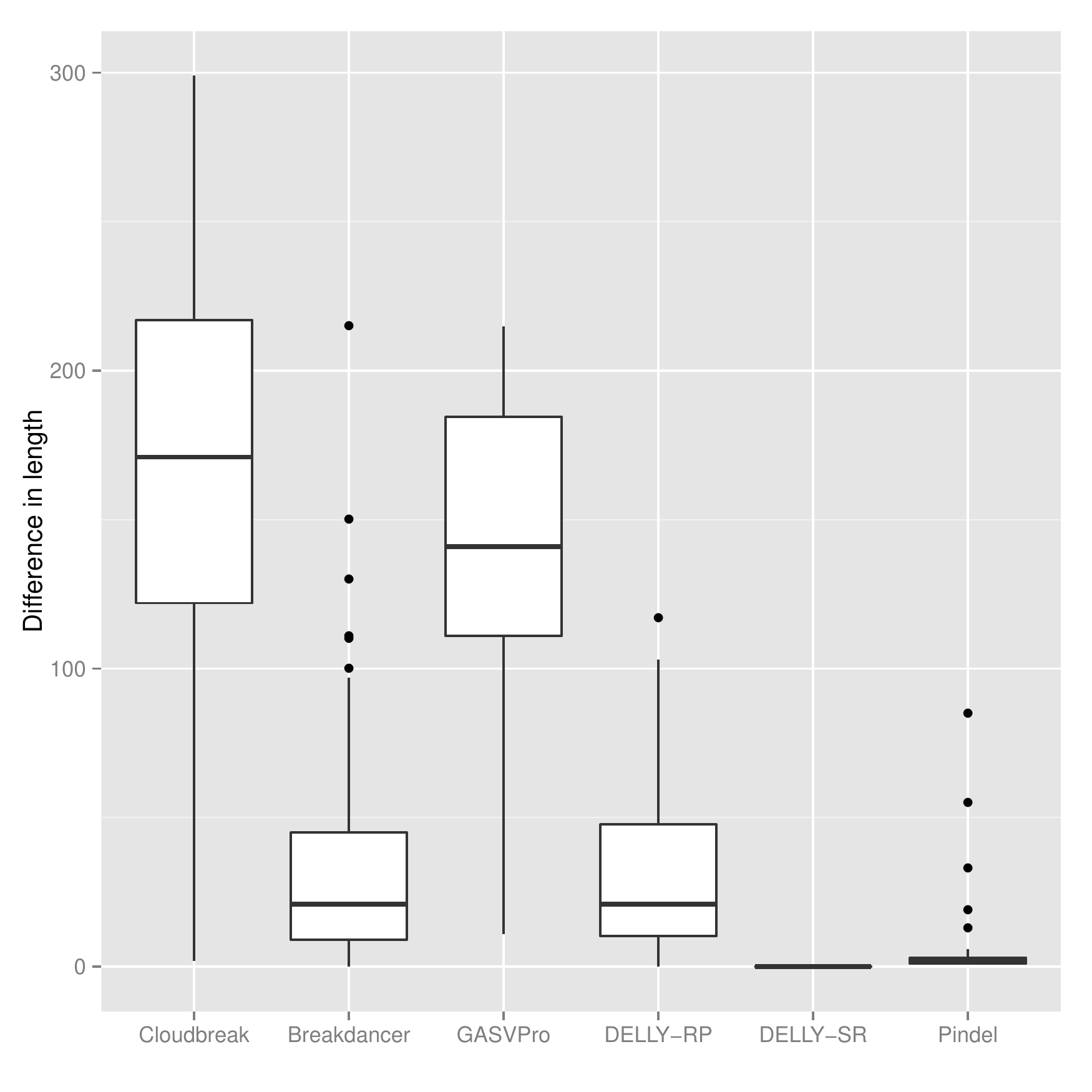}
\caption{Breakpoint resolution for each tool for deletions on the Chromosome 2 simulated data. For each correctly predicted deletion, we calculated the difference in length between the true deletion and the prediction.}
\label{breakpoint_resolution}
\end{figure}

\clearpage

\begin{suppalgorithm}[h]
\algrenewcommand\algorithmicprocedure{\textbf{job}}
  \begin{algorithmic}[1]
    \Procedure{Alignment}{}
    \Function{Map}{$\textrm{ReadPairId }rpid, \textrm{ReadId }r, \textrm{ReadSequence }s, \textrm{ReadQuality }q$}
    \ForAll{$ \textrm{Alignments }a \in \textsc{Align}(<s,q>)$}
    \State $\textsc{Emit}(\textrm{ReadPairId }rpid, \textrm{Alignment }a)$
    \EndFor
    \EndFunction
    \Function{Reduce}{$\textrm{ReadPairId }rpid, \textrm{Alignments }a_{1,2,\ldots}$}
    \State $\textrm{AlignmentPairList }ap \gets \textsc{ValidAlignmentPairs}(a_{1,2,\ldots})$
    \State $\textsc{Emit}(\textrm{ReadPairId }rp, \textrm{AlignmentPairList } ap)$
    \EndFunction
    \EndProcedure

    \Procedure{Compute SV Features}{}
    \Function{Map}{$\textrm{ReadPairId }rp, \textrm{AlignmentPairList }ap$}
    \ForAll{$ \textrm{AlignmentPairs }<a_1,a_2>  \in ap$}
    \ForAll{$ \textrm{GenomicLocations }l \in \textsc{Loci }(a_1,a_2)$}
    \State $ \textrm{ReadPairInfo }rpi \gets <\textrm{InsertSize}(a_1,a_2), \textrm{AlignmentScore}(a_1,a_2)>$
    \State $\textsc{Emit}(\textrm{GenomicLocation }l, \textrm{ReadPairInfo }rpi)$
    \EndFor
    \EndFor
    \EndFunction
    \Function{Reduce}{$\textrm{GenomicLocation }l, \textrm{ReadPairInfos }rpi_{1,2,\ldots}$}
    \State $\textrm{SVFeatures } \phi_l \gets \Phi(\textrm{InsertSizes }i_{1,2,\ldots}, \textrm{AlignmentScores }q_{1,2,\ldots})$
    \State $\textsc{Emit}(\textrm{GenomicLocation }l, \textrm{SVFeatures } \phi_l)$
    \EndFunction
    \EndProcedure

    \Procedure{Call SVs}{}
    \Function{Map}{$\textrm{GenomicLocation }l, \textrm{SVFeatures } \phi_l$}
    \State $\textsc{Emit}(\textrm{Chromosome}(l), <l,\phi_l>)$
    \EndFunction
    \Function{Reduce}{$\textrm{Chromosome }c, \textrm{GenomicLocation } l_{1,2,\ldots},\phi_{1,2,\ldots}$}
    \State $\textrm{StructuralVariationCalls } svs_c \gets \textsc{PostProcess }(\phi_{1,2,\ldots})$
    \EndFunction
    \EndProcedure
  \end{algorithmic}
\caption{The algorithmic framework for SV calling in MapReduce.}
\label{cb_algo}
\end{suppalgorithm}

\clearpage

\begin{table}[h]
\begin{center}
\begin{tabular}{r|r|rrr|rrr}
\multicolumn{2}{c}{}  & \multicolumn{3}{c}{Simulated Data} & \multicolumn{3}{c}{NA18507} \\
\hline
 & SV Types &  Single CPU & Parallel & Proc. &  Single CPU & Parallel & Proc.  \\ 
  \hline
  Cloudbreak & D,I &   NA    & 290 & 312    & NA         & 824 & 636 \\ 
  Breakdancer & D,I,V,T &  653   & NA       & NA          & 134,170 &  5,586 & 84 \\
  GASVPro & D,V   &  3,339  & NA       & NA         & 52,385  & NA & NA \\
  DELLY & D         &  1,964 & NA          & NA      & 30,311  & 20,224 & 84 \\
  Pindel & D,I,V,P         & 37,006 &  4,885     & 8          &  284,932  & 28,587 & 84 \\ 
  MoDIL & D,I        &  NA      & 52,547 & 250 & NA         & NA  & NA\\ 
   \hline
\end{tabular}
\end{center}
\caption{Runtimes (elapsed) on both data sets of each tool tested, in single-processor and parallel mode. For parallel runs, Proc. is the maximum number of simultaneously running processes or threads. All times are in seconds. The types of variants detected by each program are listed with the abbreviations: D - deletion; I - insertion; V - Inversion; P - duplication; T - translocation. Interchromosomal translocations are only detected by Breakdancer in single CPU mode. }
\label{runtimes}
\end{table}

\clearpage

\begin{table}[h]
\begin{center}
\begin{tabular}{rrrrrr}
  \hline
 & 40-100bp  & 101-250bp  & 251-500bp & 501-1000bp & $>$ 1000bp \\ 
 Total Number & 224 &  84 & 82 &  31 & 26\\ 
  \hline
  Cloudbreak  & \textbf{68} (17)  & \textbf{67} (\textbf{10}) &  \textbf{56} (\textbf{5}) & \textbf{11} (\textbf{3}) & \textbf{15} (\textbf{0}) \\ 
  Breakdancer & 52 (8)  & 49 (2) &  49 (0) & 7 (0) & 14 (\textbf{0}) \\ 
  GASVPro     & 35 (2)  & 26 (0) &  26 (0) & 2 (0) & 6 (\textbf{0}) \\ 
  DELLY-RP       & 22 (1)  & 56 (1) &  40 (0) & 8 (0) & 12 (\textbf{0}) \\ 
  DELLY-SR       & 0 (0)  & 2 (0) &  28 (0) & 2 (0) & 10 (\textbf{0}) \\ 
  Pindel      & 60 (\textbf{32})  & 16 (0) &  41 (2) & 1 (0) & 12 (\textbf{0})\\ 
   \hline
\end{tabular}
\end{center}
\caption{The number of simulated deletions in the 30X diploid chromosome 2 with Venter indels found by each tool at a 10\% FDR, as well as the number of those deletions that were discovered exclusively by each tool (in parentheses). The total number of deletions in each size class in the true set of deletions is shown in the second row of the header.}
\label{chr2DeletionPredsFDR10}
\end{table}

\newpage 

\begin{table}[h]
\begin{center}
\begin{tabular}{rrr|rr}
 & \multicolumn{2}{c}{Simulated Data} & \multicolumn{2}{c}{NA18507} \\
\hline
 &  Non-repeat & Repeat  &  Non-repeat & Repeat \\ 
 Total Number & 120 & 327 & 562 & 8059 \\ 
  \hline
  Cloudbreak  & \textbf{37} (\textbf{10}) & \textbf{180} (\textbf{25}) & \textbf{300} (\textbf{85}) & \textbf{1166} (\textbf{270}) \\ 
  Breakdancer & 29 (7) & 142 (3) & 192 (11) & 868 (21) \\
  GASVPro     & 16 (1) & 79 (1) & 79 (7) & 330 (16) \\
  DELLY-RP       & 21 (1) & 117 (1) & 152 (6) & 712 (5) \\
  DELLY-SR       & 0 (0) & 42 (0) & 27 (0) & 391 (0) \\
  Pindel      & 18 (9) & 112 (\textbf{25}) & 109 (2) & 536 (7) \\ 
   \hline
\end{tabular}
\end{center}
\caption{Detected deletions on the simulated and NA18507 data sets identified by each tool, broken down by whether the deletion overlaps with a RepeatMasker-annotated element. For both data sets we used the thresholds derived from finding the 10\% FDR level in the simulated data set. The proportion of deletions that overlap repetitive elements discovered by Cloudbreak is similar to that of the other methods.}
\label{deletionRepmaskpreds}
\end{table}

\newpage

\begin{table}[h]
\begin{center}
\begin{tabular}{rrr|rr}
 & \multicolumn{2}{c}{Simulated Data} & \multicolumn{2}{c}{NA18507} \\
\hline
 &  Non-repeat & Repeat  &  Non-repeat & Repeat \\ 
 Total Number & 133 & 270 & 341 & 355 \\ 
  \hline
  Cloudbreak  & \textbf{32} (11) & \textbf{91} (\textbf{47}) & \textbf{169} (\textbf{61}) & \textbf{148} (\textbf{68}) \\ 
  Breakdancer & 17 (5) & 22 (6) & 82 (7) & 44 (9) \\
  Pindel      & 25 (\textbf{16}) & 54 (30) & 84 (35) & 82 (24) \\ 
  MoDIL      & 5 (0) & 16 (3) & NA & NA \\ 
   \hline
\end{tabular}
\end{center}
\caption{Detected insertions in the simulated and NA18507 data sets identified by each tool, broken down by whether the insertion occurs in a RepeatMasker-annotated element. The maximum sensitivity cutoffs were used for both data sets. The proportion of insertions in repetitive elements discovered by Cloudbreak is similar to that of the other methods.}
\label{insertionRepmaskpreds}
\end{table}

\newpage

\begin{table}[h]
\begin{center}
\begin{tabular}{r|r|rr|rr|}
\multicolumn{2}{c}{}  & \multicolumn{4}{c}{Actual Genotypes} \\
\multicolumn{2}{c}{}  & \multicolumn{2}{c}{Simulated Data} & \multicolumn{2}{c}{NA18507} \\
\cline{3-6}
\multicolumn{2}{c|}{} &  Homozygous & Heterozygous & Homozygous & Heterozygous \\ 
\cline{2-6}
\multirow{2}{*}{\shortstack{Predicted \\ Genotypes}} & Homozygous & 35 & 2 &  96 & 21 \\
 & Heterozygous & 0 & 39 &  2 & 448 \\
\cline{2-6}
\end{tabular}
\end{center}
\caption{Confusion matrices for the predicted genotype of deletions found by Cloudbreak on both the simulated and NA18507 data sets.}
\label{deletionGenotypeaccuracy}
\end{table}

\newpage 

\section{Supplementary Discussion}

\subsection{A general framework for SV Detection in MapReduce}

We have developed a conceptual algorithmic framework for SV detection in MapReduce, which is outlined in Supplementary Algorithm \ref{cb_algo}. As described in the main text, this framework divides processing into three separate MapReduce jobs: an alignment job, a feature computation job, and an SV calling job. 

The alignment job uses sensitive mapping tools and settings to discover mapping locations for each read pair. Aligners can be executed to report multiple possible mappings for each read, or only the best possible mapping. Given a set of read pairs, each of which consists of a read pair identifier $rpid$ and two sets of sequence and quality scores $<s,q>$, each mapper aligns each pair end set $<s,q>$ in either single- or paired end mode and emits possible mapping locations under the $rpid$ key. Reducers then collect the alignments for each paired end, making them available under one key for the next job. Our implementation contains wrappers to execute the aligners BWA, GEM \autocite{MarcoSola:2012hm}, Novoalign\footnote{\url{http://wwww.novocraft.com}}, RazerS 3 \autocite{Weese:2012by}, mrFAST \autocite{Alkan:2009cr}, and Bowtie 2 \autocite{Langmead:2012jh}, as well as the ability to import a pre-aligned BAM file directly to HDFS.

In the second job, we compute a set of features for each location in the genome. To begin, we tile the genome with small fixed-width, non-overlapping intervals. For the experiments reported in this paper we use an interval size of 25bp. Let $L = \left\{l_1,l_2,\ldots,l_N\right\}$ be the set of intervals covering the entire genome. Let $R^1 = \left\{r^{1}_{1},r^{1}_{2},\ldots,r^{1}_{M}\right\}$ and $R^2 = \left\{r^{2}_{1},r^{2}_{2},\ldots,r^{2}_{M}\right\}$ be the input set of paired reads. Let $A^1 = \left\{a^{1}_{m,1},a^{1}_{m,2},\ldots,a^{1}_{m,K}\right\}$ and $A^2 = \left\{a^{2}_{m,1},a^{2}_{m,2},\ldots,a^{2}_{m,L}\right\}$ be the set of alignments for the left and right reads from read pair $m$. For any given pair of alignments of the two reads in a read pair, $a^{1}_{m,i}$ and $a^{2}_{m,j}$, let the $\textrm{ReadPairInfo } rpi_{m,i,j}$ be information about the pair relevant to detecting SVs, e.g. the fragment size implied by the alignments and the likelihood the alignments are correct. We then leave two functions to be implemented depending on the application:
\begin{flalign*}
 \textsc{Loci } :& \langle a^{1}_{m,i},a^{2}_{m,j} \rangle \rightarrow L_m \subseteq L \\
 \Phi :& \left\{\textrm{ReadPairInfo }rpi_{m,i,j}\right\} \rightarrow \mathbb{R}^N \\
\end{flalign*}

The first function, \textsc{Loci}, maps an alignment pair to a set of genomic locations to which it is relevant for SV detection; for example, the set of locations overlapped by the internal insert implied by the read alignments. We optimize this step by assuming that if there exist concordant mappings for a read pair, defined as those where the two alignments are in the proper orientation and with an insert size within three standard deviations of the expected library insert size, one of them is likely to be correct and therefore we do not consider any discordant alignments of the pair. The second function, $\Phi$, maps a set of ReadPairInfos relevant to a given location to a set of real-valued vectors of features useful for SV detection. 

Finally, the third MapReduce job is responsible for making SV calls based on the features computed at each genomic location. It calls another application-specific function  $\textsc{PostProcess} : \left\{\phi_1,\phi_2,\ldots,\phi_N\right\} \rightarrow \left\{\langle  \textrm{SVType } s, l_{start}, l_{end} \rangle\right\}$  that maps the sets of features for related loci into a set of SV calls characterized by their type $s$ (i.e Deletion, Insertion, etc.) and their breakpoint locations $l_{start}$ and $l_{end}$. We parallelize this job in MapReduce by making calls for each chromosome in parallel, which we achieve by associating a location and its set of features to its chromosome in the map phase, and then making SV calls for one chromosome in each reduce task.

\subsection{Cloudbreak: An Implementation of the MapReduce SV Detection Framework}

Our Cloudbreak software package can be seen as an implementation of the general framework defined above. In particular, Cloudbreak implements the three user-defined function described above as follows:

\begin{description}
\item[\sc{Loci}] Because we are detecting deletions and short insertions, we map ReadPairInfos from each possible alignment to the genomic locations overlapped by the implied internal insert between the reads. For efficiency, we define a maximum detectable deletion size of 25,000bp, and therefore alignment pairs in which the ends are more than 25kb apart, or in the incorrect orientation, map to no genomic locations.
\item[$\Phi$] To compute features for each genomic location, we follow \textcite{Lee:2009da}, who observed that if all mappings are correct, the insert sizes implied by mappings which span a given genomic location should follow a Gaussian mixture model (GMM) whose parameters depend on whether a deletion or insertion is present at that locus (Supplementary Figure \ref{insert_size_mixes}). Briefly: if there is no indel, the insert sizes implied by spanning alignment pairs should follow the distribution of actual fragment sizes in the sample, which is typically modeled as normally distributed with mean $\mu$ and standard deviation $\sigma$. If there is a homozygous deletion or insertion of length $l$ at the location, $\mu$ should be shifted to $\mu + l$, while $\sigma$ will remain constant. Finally, in the case of a heterozygous event, the distribution of insert sizes will follow a mixture of two normal distributions, one with mean $\mu$, and the other with mean $\mu + l$, both with an unchanged standard deviation of $\sigma$, and mixing parameter $\alpha$ that describes the relative weights of the two components. Because the mean and standard deviation of the fragment sizes are selected by the experimenter and therefore known \emph{a priori} (or at least easily estimated based on a sample of alignments), we only need to estimate the mean of the second component at each locus, and the mixing parameter $\alpha$.

To handle incorrect and ambiguous mappings, we assume that in general they will not form normally distributed clusters in the same way that correct mappings will, and therefore use an outlier detection technique to filter the observed insert sizes for each location. We sort the observed insert sizes and define as an outlier an observation whose $k$th nearest neighbor is more than $n\sigma$ distant, where $k = 3$ and $n = 5$. In addition, we rank all observations by the estimated probability that the mapping is correct and use an \emph{adaptive quality cutoff} to filter observations: we discard all observations where the estimated probability the mapping is correct is less than the score of the maximum quality observation minus a constant $c$. This allows the use of more uncertain mappings in repetitive regions of the genome while restricting the use of low-quality mappings in unique regions. Defining $\textsc{Mismatches}(a)$ to be the number of mismatches between a read and the reference genome in the alignment $a$, we approximate the probability $p^{k}_c$ of each end alignment being correct by:

\[ p^{k}_c(a^{k}_{m,i}) = \frac{\exp({-\textsc{Mismatches}(a^{k}_{m,i})/2)}}{\sum_j{\exp(-\textsc{Mismatches}(a^{k}_{m,j})/2)}} \]

And then multiply $p_c(a^{1}_{m,i})$ and $p_c(a^{2}_{m,i})$ to approximate the likelihood that the pair is mapped correctly.

We fit the parameters of the GMM using the Expectation-Maximization algorithm. Let $Y = y_{1,2, \ldots m}$ be the observed insert sizes at each location after filtering, and say the library has mean fragment size $\mu$ with standard deviation $\sigma$. We initialize the two components to have means $\mu$ and $\bar{Y}$, set the standard deviation of both components to $\sigma$, and set $\alpha = .5$. In the E step, we compute for each $y_i$ and GMM component $j$ the value $\gamma_{i,j}$, which is the normalized likelihood that $y_i$ was drawn from component $j$. We also compute $n_j = \sum_i{\gamma_{i,j}}$, the relative contributions of the data points to each of the two distributions. In the M step, we update $\alpha$ to be $n_2 - \left|Y\right|$, and set the mean of the second component to be $\frac{\sum_m{\gamma_{m,2}y_m}}{n_2}$. We treat the variance as fixed and do not update it, since under our assumptions the standard deviation of each component should always be $\sigma$. We repeat the E and M steps until convergence, or until a maximum number of steps has been taken.

The features generated for each location $l$ include the log-likelihood ratio of the filtered observed data points under the fit GMM to their likelihood under the distribution $N(\mu,\sigma)$, the final value of the mixing parameter $\alpha$, and $\mu'$, the estimated mean of the second GMM component.

\item[\sc{PostProcess}] We convert our features along the genome to insertion and deletion calls by first extracting contiguous genomic loci where the log-likelihood ratio of the two models is greater than a given threshold. To eliminate noise we apply a median filter with window size 5. We end regions when $\mu'$ changes by more than 60bp ($2\sigma$), and discard regions where the average value of $\mu'$ is less than $\mu$ or where the length of the region differs from $\mu'$ by more than $\mu$.
\end{description}

\subsection{Runtime Analysis}

We implemented and executed Cloudbreak on a 56-node Hadoop cluster, with 636 map slots and 477 reduce slots. Not including alignment time, we were able to process the Chromosome 2 simulated data in under five minutes, and the the NA18507 data set in under 15 minutes. For the simulated data set we used 100 reducers for the compute SV features job; for the real data set we used 300. The bulk of Cloudbreak's execution is spent in the feature generation step. Extracting deletion and insertion calls take under two minutes each for both the real and simulated data sets; the times are equal because each reducer is responsible for processing a single chromosome, and so the runtime is bounded by the length of time it takes to process the largest chromosome. 

In Supplementary Table \ref{runtimes} we display a comparison of runtimes on the real and simulated data sets for all of the tools evaluated in this work. Each tool varies in the amount of parallelization supported. We report runtimes for tools run in their default single-threaded mode, as well as for levels of parallelization achievable with basic scripting, noting that one of the key advantages of Hadoop/MapReduce is the ability to scale parallel execution to the size of the available compute cluster without any custom programming. Pindel allows multi-threaded operation on multicore servers. Pindel and Breakdancer allow processing of a single chromosome in one process, so it is possible to execute all chromosomes in parallel on a cluster that has a job scheduler and shared filesystem. Breakdancer has an additional preprocessing step (\texttt{bam2cfg.pl}) which runs in a single thread. DELLY suggests splitting the input BAM file by chromosome, after which a separate DELLY process can be executed on the data for each chromosome; splitting a large BAM file is a time consuming process and consumes most of the time in this parallel workflow, in fact making it faster to run in single-threaded mode. GASVPro allows parallelization of the MCMC component for resolving ambiguously mapped read pairs; however, this requires a significant amount of custom scripting, and we did not find that the MCMC module consumed most of the runtime in our experiments, so we do not attempt to parallelize this component. The MoDIL distribution contains a set of scripts that can be used to submit parallel jobs to the SGE scheduling engine or modified for other schedulers; we adapted these for use in our cluster.

In parallel execution, the total time to execute is bounded by the runtime of the longest-running process. In the case of chromosome-parallelizable tools including Breakdancer, Pindel, and DELLY, this is typically the process working on the largest chromosome.\footnote{We note that one Breakdancer process, handling an unplaced contig in the hg19 reference genome, never completed in our runs and had to be killed manually; we exclude that process from our results.} In the case of MoDIL's run on the simulated data, we found that the different processes varied widely in their execution times, likely caused by regions of high coverage or with many ambiguously mapped reads. Cloudbreak mitigates this problem during the time-consuming feature generation process by using Hadoop partitioners to randomly assign each genomic location to one of the set of reducers, ensuring that the work is evenly distributed across all processes. This distribution of processing across the entire cluster also serves to protect against server slowdowns and hardware failures - for example, we were still able to complete processing of the NA18507 data set during a run where one of the compute nodes was rebooted midway through the feature generation job.

\printbibliography[title={Supplementary References}]

\newpage

\section{Cloudbreak User Manual}
\label{cloudbreak}

Cloudbreak is a Hadoop-based structural variation (SV) caller for Illumina
paired-end DNA sequencing data. Currently Cloudbreak calls genomic insertions
and deletions; we are working on adding support for other types of SVs.

Cloudbreak contains a full pipeline for aligning your data in the form of FASTQ
files using alignment pipelines that generate many possible mappings for every
read, in the Hadoop framework. It then contains Hadoop jobs for computing
genomic features from the alignments, and for calling insertion and deletion
variants from those features.

You can get Cloudbreak by downloading a pre-packaged release from the ``releases''
tab in the GitHub repository, or by building from source as described below.

\subsection{Building From Source}
\label{buildingfromsource}

To build the latest version of Cloudbreak, clone the GitHub repository. You'll
need to install Maven to build the executables. (http:/\slash maven.apache.org\slash )
Enter the top level directory of the Cloudbreak repository and type the command:

\texttt{mvn package}

This should compile the code, execute tests, and create a distribution file, \linebreak
 \texttt{cloudbreak-\$VERSION-dist.tar.gz}, in the \texttt{target\slash } directory. You can then copy
 that distribution file to somewhere else on your system, unpack it with:

 \texttt{tar -xzvf cloudbreak-\$VERSION-dist.tar.gz} 

and access the Cloudbreak jar file and related scripts and properties files.

\subsection{Dependencies}
\label{dependencies}

Cloudbreak requires a cluster Hadoop 0.20.2 or Cloudera CDH3 to run (the older
mapreduce API). If you don't have a Hadoop cluster, Cloudbreak can also use the
Apache Whirr API to automatically provision a cluster on the Amazon Elastic
Compute Cloud (EC2). See the section on using WHIRR below.

If you wish to run alignments using Cloudbreak, you will need one of the following
supported aligners:

\begin{itemize}
\item BWA (Recommended): http:/\slash bio-bwa.sourceforge.net\slash 

\item GEM: http:/\slash algorithms.cnag.cat\slash wiki\slash The\_GEM\_library

\item RazerS 3: http:/\slash www.seqan.de\slash projects\slash razers\slash 

\item Bowtie2: http:/\slash bowtie-bio.sourceforge.net\slash bowtie2\slash index.shtml

\item Novoalign: http:/\slash www.novocraft.com

\end{itemize}

\subsection{User Guide}
\label{userguide}

You can use Cloudbreak in several different ways, depending on whether you want
to start with FASTQ files and use Hadoop to help parallelize your alignments, or if you already
have an aligned BAM file and just want to use Cloudbreak to call variants. In addition,
the workflow is slightly different depending on whether you want to run on a local
Hadoop cluster or want to run using a cloud provider like Amazon EC2. Later in this
file, we've listed a set of scenarios to describe options for running the Cloudbreak
pipeline. Find the scenario that best fits your use case for more details on how to
run that workflow. For each scenario, we have created a template script that contains
all of the steps and parameters you need, which you can modify for your particular data set.

\subsection{Running on a cloud provider like Amazon EC2 with Whirr}
\label{runningonacloudproviderlikeamazonec2withwhirr}

Cloudbreak has support for automatically deploying a Hadoop cluster on
cloud providers such as Amazon EC2, transferring your data there, running the Cloudbreak algorithm, and
downloading the results.

Of course, renting compute time on EC2 or other clouds costs money, so please be
familiar with the appropriate usage and billing policies of your cloud provider
before attempting this.

WE ARE NOT RESPONSIBLE FOR UNEXPECTED CHARGES THAT YOU MIGHT INCUR ON EC2 OR
OTHER CLOUD PROVIDERS.

Many properties that affect the cluster created can be set in the file
\texttt{cloudbreak-whirr.properties} in this distribution. You will need to edit this file
to set your AWS access key and secret access key (or your credentials for other
cloud provider services), and to tell it the location of the public and
private SSH keys to use to access the cluster. You can also control the number
and type of nodes to include in the cluster. The default settings in the file
create 15 nodes of type m1.xlarge, which is sufficient to fully process a 30X
simulation of human chromosome 2, including read alignment and data transfer time,
in under an hour. We have only tested this capability using EC2; other cloud providers
may not work as well. You can also direct Whirr to use Amazon EC2's spot instances, which are
dramatically cheaper than on-demand instances, although they carry the risk of
being terminated if your price is out-bid. Using recent spot pricing, it cost
us about \$5 to run the aforementioned chromosome 2 simulation. We recommend
setting your spot bid price to be the on demand price for the instance type you
are using to minimize the chance of having your instances terminated.

Please consult Amazon's EC2 documentation and the documentation for Whirr for
more information on how to configure and deploy clusters in the cloud.

\subsection{Running on a Small Example Data Set}
\label{runningonasmallexampledataset}

To facilitate testing of Cloudbreak, we have publicly hosted the reads from the simulated
data example described in the Cloudbreak manuscript on a bucket in Amazon's S3 storage
 service at s3:/\slash cloudbreak-example\slash . We have also provided an example script that creates
 a cluster in Amazon EC2, copies the data to the cluster, runs the full Cloudbreak
 workflow including alignments with BWA, and copies the variant calls back to the
 local machine before destroying the cluster. The script, called \texttt{Cloudbreak-EC2-whirr-example.sh}
 is in the scripts directory of the Cloudbreak distribution. Of course, you will still
 need to edit the \texttt{cloudbreak-whirr.properties} file with your EC2 credentials, and verify
 that the cluster size, instance types, and spot price are to your liking before
 executing the example.

\subsubsection{Scenario 1: Compute alignments in Hadoop, using a local Hadoop cluster}
\label{scenario1:computealignmentsinhadoopusingalocalhadoopcluster}

To install aligner dependencies for use by Cloudbreak, first generate the index
for the genome reference you would like to run against. Then, copy all of the
required index files, and the executable files for the aligner into HDFS using
the \texttt{hadoop dfs -copyFromLocal} command. For BWA you will need all of the index files
created by running \texttt{bwa index}. You will also need an `fai' file for the reference,
containing chromosome names and lengths, generated by \texttt{samtools faidx}.

If your reference file is \texttt{reference.fa}, and \texttt{bwa aln} has created the files

\begin{verbatim}
reference.fa.amb
reference.fa.ann
reference.fa.bwt
reference.fa.pac
reference.fa.sa
\end{verbatim}

and \texttt{reference.fa.fai} as described above, issue the following
commands to load the necessary files into HDFS:

\begin{verbatim}
hdfs -mkdir indices/
hdfs -mkdir executables/
hdfs -copyFromLocal reference.fa.amb indices/
hdfs -copyFromLocal reference.fa.ann indices/
hdfs -copyFromLocal reference.fa.bwt indices/
hdfs -copyFromLocal reference.fa.pac indices/
hdfs -copyFromLocal reference.fa.sa indices/
hdfs -copyFromLocal reference.fa.fai indices/
hdfs -copyFromLocal /path/to/bwa/executables/bwa executables/
hdfs -copyFromLocal /path/to/bwa/executables/xa2multi.pl executables/
\end{verbatim}

The basic workflow is:

\begin{enumerate}
\item Load the FASTQ files into HDFS

\item Run one of the Cloudbreak alignment commands to align your reads

\item Create a readGroup file to describe the location and insert size characteristics of your reads, and copy it into HDFS.

\item Run the GMM fitting feature generation step of the Cloudbreak process.

\item Extract deletion calls from the features created in step 4.

\item Copy the deletion calls from HDFS to a local directory.

\item Extract insertion calls from the features created in step 4.

\item Copy the insertion calls from HDFS to a local directory.

\item Optionally, export the alignments back into a BAM file in your local filesystem.

\end{enumerate}

We have created a script to run through the full process of executing the
Cloudbreak pipeline from FASTQ files to insertion and deletion calls. The
script is named \texttt{Cloudbreak-full.sh} and can be found in the scripts directory
of the Cloudbreak distribution. To customize the script for your needs, copy
it to a new location and edit the variables in the first three sections:
``EXPERIMENT DETAILS'', ``LOCAL FILES AND DIRECTORIES'', and
``HDFS FILES AND DIRECTORIES''.

\subsubsection{Scenario 2: Call variants on existing alignments, using a local Hadoop cluster}
\label{scenario2:callvariantsonexistingalignmentsusingalocalhadoopcluster}

For this scenario you don't need to worry about having an aligner executable or
aligner-generated reference in HDFS. You will however, need a chromosome length
`fai' file, which you can generate by running \texttt{samtools faidx} on your reference
FASTA files and then copying to HDFS:

\begin{verbatim}
hdfs -copyFromLocal reference.fa.fai indices/
\end{verbatim}

After that, the workflow is:

\begin{enumerate}
\item Load your BAM file into HDFS and prepare it for Cloudbreak

\item Create a readGroup file to describe the location and insert size characteristics of your reads.

\item Run the GMM fitting feature generation step of the Cloudbreak process.

\item Extract deletion calls from the features created in step 3.

\item Copy the deletion calls from HDFS to a local directory.

\item Extract insertion calls from the features created in step 3.

\item Copy the insertion calls from HDFS to a local directory.

\end{enumerate}

To prepare alignments for Cloudbreak, they must be sorted by read name. You can then use the
 \texttt{readSAMFileIntoHDFS} Cloudbreak command.

A templates for this scenario is available in the script \texttt{Cloudbreak-variants-only.sh}
located in the scripts directory of the Cloudbreak distribution.

\subsubsection{Scenario 3: Compute alignments in Hadoop, using a cloud provider like EC2}
\label{scenario3:computealignmentsinhadoopusingacloudproviderlikeec2}

First, see the section ``Running on a Cloud Provider like Amazon EC2 with Whirr'' above, and modify the file
\texttt{cloudbreak-whirr.properties} to include your access credentials and the appropriate cluster
specifications. After that, the workflow is similar to the workflow described for scenario \#1
above, with the additional first steps of copying your reads and dependency files to the cloud and
creating a cluster before processing begins, and then destroying the cluster after processing has
completed.

You can see an example workflow involving EC2 by examining the script
\texttt{Cloudbreak-EC2-whirr.sh}. This begins by transferring your reads to Amazon S3. It then
uses Apache Whirr to launch an EC2 Hadoop cluster, copies the necessary executable files
to EC2, and runs the algorithm.

\subsubsection{Scenario 4: Call variants on existing alignments, using a cloud provider like EC2}
\label{scenario4:callvariantsonexistingalignmentsusingacloudproviderlikeec2}

Again, please read the section ``Running on a Cloud Provider like Amazon EC2 with Whirr'' above to learn how to
update the \texttt{cloudbreak-whirr.properties} file with your credentials and cluster specifications. After that,
follow the template in the script \texttt{Cloudbreak-EC2-whirr-variants-only.sh} to create a workflow
involving calling variants in the cloud.

\subsection{Output Files}
\label{outputfiles}

The output from running Cloudbreak using one of the scripts above will be found in the files named

\begin{verbatim}
READ_GROUP_LIBRARY_dels_genotyped.bed
READ_GROUP_LIBRARY_ins_genotyped.bed
\end{verbatim}

where READ\_GROUP and LIBRARY are the names of the reads in your experiment. The
format of the files is tab-delimited with the following columns:

\begin{itemize}
\item CHROMOSOME: The chromosome of the deletion call

\item START: The start coordinate of the deletion call

\item END: The end coordinate of the deletion call

\item NUMBER: The cloudbreak identifier of the deletion call

\item LR: The likelihood ratio of the deletion (higher indicates a call more likely to be true)

\item TYPE: Either ``INS'' or ``DEL''

\item W: The average weight of the estimated GMM mixing parameter alpha, used in genotyping

\item GENOTYPE: The predicted genotype of the call

\end{itemize}

\subsection{Contact information}
\label{contactinformation}

Please contact cwhelan at gmail.com with any questions on running cloudbreak.

\subsection{Reference Guide}
\label{referenceguide}

All of Cloudbreak's functionality is contained in the executable jar file in the
directory where you unpacked the Cloudbreak distribution. Use the `hadoop'
command to run the jar file to ensure that the necessary Hadoop dependencies
are available to Cloudbreak.

To invoke any Cloudbreak command, use a command line in this format:

\texttt{hadoop cloudbreak-\$\{project.version\}.jar [options] [command] [command options]}

Where \texttt{command} is the name of the command, \texttt{command options} are the arguments specific
to that command, and \texttt{options} are general options, including options for how to run
Hadoop jobs. For example, if you'd like to specify 50 reduce tasks
for one of your commands, pass in \texttt{-Dmapred.reduce.tasks=50} as one of the general options. 

Each command is detailed below and its options are listed below. You can view this information by typing
\texttt{hadoop jar cloudbreak-\$\{project.version\}.jar} without any additional parameters.

\begin{verbatim}
    readPairedEndFilesIntoHDFS      Load paired FASTQ files into HDFS
      Usage: readPairedEndFilesIntoHDFS [options]
  Options:
        *     --HDFSDataDir                  HDFS directory to load reads into
              --clipReadIdsAtWhitespace      Whether to clip all readnames at
                                             the first whitespace (prevents trouble
                                             with some aligners)
                                             Default: true
              --compress                     Compression codec to use on the
                                             reads stored in HDFS
                                             Default: snappy
        *     --fastqFile1                   File containing the first read in
                                             each pair
        *     --fastqFile2                   File containing the second read in
                                             each pair
              --filesInHDFS                  Use this flag if the BAM file has
                                             already been copied into HDFS
                                             Default: false
              --filterBasedOnCasava18Flags   Use the CASAVA 1.8 QC filter to
                                             filter out read pairs
                                             Default: false
              --outFileName                  Filename of the prepped reads in
                                             HDFS
                                             Default: reads
              --trigramEntropyFilter         Filter out read pairs where at
                                             least one read has a trigram entropy less
                                             than this value. -1 = no filter
                                             Default: -1.0

    readSAMFileIntoHDFS      Load a SAM/BAM file into HDFS
      Usage: readSAMFileIntoHDFS [options]
  Options:
        *     --HDFSDataDir   HDFS Directory to hold the alignment data
              --compress      Compression codec to use for the data
                              Default: snappy
              --outFileName   Filename to give the file in HDFS
                              Default: alignments
        *     --samFile       Path to the SAM/BAM file on the local filesystem

    bwaPairedEnds      Run a BWA paired-end alignment
      Usage: bwaPairedEnds [options]
  Options:
        *     --HDFSAlignmentsDir    HDFS directory to hold the alignment data
        *     --HDFSDataDir          HDFS directory that holds the read data
        *     --HDFSPathToBWA        HDFS path to the bwa executable
              --HDFSPathToXA2multi   HDFS path to the bwa xa2multi.pl executable
        *     --maxProcessesOnNode   Ensure that only a max of this many BWA
                                     processes are running on each node at once.
                                     Default: 6
              --numExtraReports      If > 0, set -n and -N params to bwa sampe,
                                     and use xa2multi.pl to report multiple hits
                                     Default: 0
        *     --referenceBasename    HDFS path of the FASTA file from which the
                                     BWA index files were generated.

    novoalignSingleEnds      Run a Novoalign alignment in single ended mode
      Usage: novoalignSingleEnds [options]
  Options:
        *     --HDFSAlignmentsDir            HDFS directory to hold the
                                             alignment data
        *     --HDFSDataDir                  HDFS directory that holds the read
                                             data
        *     --HDFSPathToNovoalign          HDFS path to the Novoalign
                                             executable
              --HDFSPathToNovoalignLicense   HDFS path to the Novoalign license
                                             filez
              --qualityFormat                Quality score format of the FASTQ
                                             files
                                             Default: ILMFQ
        *     --reference                    HDFS path to the Novoalign
                                             reference index file
        *     --threshold                    Quality threshold to use for the -t
                                             parameter

    bowtie2SingleEnds      Run a bowtie2 alignment in single ended mode
      Usage: bowtie2SingleEnds [options]
  Options:
        *     --HDFSAlignmentsDir       HDFS directory to hold the alignment
                                        data
        *     --HDFSDataDir             HDFS directory that holds the read data
        *     --HDFSPathToBowtieAlign   HDFS path to the bowtie2 executable
        *     --numReports              Max number of alignment hits to report
                                        with the -k option
        *     --reference               HDFS path to the bowtie 2 fasta
                                        reference file

    gemSingleEnds      Run a GEM alignment
      Usage: gemSingleEnds [options]
  Options:
        *     --HDFSAlignmentsDir     HDFS directory to hold the alignment data
        *     --HDFSDataDir           HDFS directory that holds the read data
        *     --HDFSPathToGEM2SAM     HDFS path to the gem-2-sam executable
        *     --HDFSPathToGEMMapper   HDFS path to the gem-mapper executable
        *     --editDistance          Edit distance parameter (-e) to use in the
                                      GEM mapping
                                      Default: 0
        *     --maxProcessesOnNode    Maximum number of GEM mapping processes to
                                      run on one node simultaneously
                                      Default: 6
        *     --numReports            Max number of hits to report from GEM
        *     --reference             HDFS path to the GEM reference file
              --strata                Strata parameter (-s) to use in the GEM
                                      mapping
                                      Default: all

    razerS3SingleEnds      Run a razerS3 alignment
      Usage: razerS3SingleEnds [options]
  Options:
        *     --HDFSAlignmentsDir   HDFS directory to hold the alignment data
        *     --HDFSDataDir         HDFS directory that holds the read data
        *     --HDFSPathToRazerS3   HDFS path to the razers3 executable file
        *     --numReports          Max number of alignments to report for each
                                    read
        *     --pctIdentity         RazerS 3 percent identity parameter (-i)
                                    Default: 0
        *     --reference           HDFS path to the reference (FASTA) file for
                                    the RazerS 3 mapper
        *     --sensitivity         RazerS 3 sensitivity parameter (-rr)
                                    Default: 0

    mrfastSingleEnds      Run a novoalign mate pair alignment
      Usage: mrfastSingleEnds [options]
  Options:
        *     --HDFSAlignmentsDir   HDFS directory to hold the alignment data
        *     --HDFSDataDir         HDFS directory that holds the read data
        *     --HDFSPathToMrfast    HDFS path to the mrfast executable file
        *     --reference           HDFS path to the mrfast reference index file
              --threshold           MrFAST threshold parameter (-e)
                                    Default: -1

    exportAlignmentsFromHDFS      Export alignments in SAM format
      Usage: exportAlignmentsFromHDFS [options]
  Options:
              --aligner        Format of the alignment records
                               (sam|mrfast|novoalign)
                               Default: sam
        *     --inputHDFSDir   HDFS path to the directory holding the alignment
                               reccords

    GMMFitSingleEndInsertSizes      Compute GMM features in each bin across the genome
      Usage: GMMFitSingleEndInsertSizes [options]
  Options:
              --aligner                            Format of the alignment
                                                   records (sam|mrfast|novoalign)
                                                   Default: sam
              --chrFilter                          If filter params are used,
                                                   only consider alignments in the
                                                   region
                                                   chrFilter:startFilter-endFilter
              --endFilter                          See chrFilter
              --excludePairsMappingIn              HDFS path to a BED file. Any
                                                   reads mapped within those intervals
                                                   will be excluded from the
                                                   processing
        *     --faidx                              HDFS path to the chromosome
                                                   length file for the reference genome
        *     --inputFileDescriptor                HDFS path to the directory
                                                   that holds the alignment records
              --legacyAlignments                   Use data generated with an
                                                   older version of Cloudbreak
                                                   Default: false
              --mapabilityWeighting                HDFS path to a BigWig file
                                                   containing genome uniqness scores. If
                                                   specified, Cloudbreak will weight reads
                                                   by the uniqueness of the regions
                                                   they mapped to
              --maxInsertSize                      Maximum insert size to
                                                   consider (= max size of deletion
                                                   detectable)
                                                   Default: 25000
              --maxLogMapqDiff                     Adaptive quality score cutoff
                                                   Default: 5.0
              --maxMismatches                      Max number of mismatches
                                                   allowed in an alignment; all other
                                                   will be ignored
                                                   Default: -1
              --minCleanCoverage                   Minimum number of spanning
                                                   read pairs for a bin to run the
                                                   GMM fitting procedure
                                                   Default: 3
              --minScore                           Minimum alignment score (SAM
                                                   tag AS); all reads with lower AS
                                                   will be ignored
                                                   Default: -1
        *     --outputHDFSDir                      HDFS path to the directory
                                                   that will hold the output of the
                                                   GMM procedure
              --resolution                         Size of the bins to tile the
                                                   genome with
                                                   Default: 25
              --startFilter                        See chrFilter
              --stripChromosomeNamesAtWhitespace   Clip chromosome names from
                                                   the reference at the first
                                                   whitespace so they match with alignment
                                                   fields
                                                   Default: false

    extractDeletionCalls      Extract deletion calls into a BED file
      Usage: extractDeletionCalls [options]
  Options:
        *     --faidx                Chromosome length file for the reference
        *     --inputHDFSDir         HDFS path to the GMM fit feature results
              --medianFilterWindow   Use a median filter of this size to clean
                                     up the results
                                     Default: 5
        *     --outputHDFSDir        HDFS Directory to store the variant calls
                                     in
              --resolution           Size of the bins to tile the genome with
                                     Default: 25
        *     --targetIsize          Mean insert size of the library
                                     Default: 0
        *     --targetIsizeSD        Standard deviation of the insert size of
                                     the library
                                     Default: 0
              --threshold            Likelihood ratio threshold to call a
                                     variant
                                     Default: 1.68

    extractInsertionCalls      Extract insertion calls into a BED file
      Usage: extractInsertionCalls [options]
  Options:
        *     --faidx                Chromosome length file for the reference
        *     --inputHDFSDir         HDFS path to the GMM fit feature results
              --medianFilterWindow   Use a median filter of this size to clean
                                     up the results
                                     Default: 5
              --noCovFilter          filter out calls next to a bin with no
                                     coverage - recommend on for BWA alignments, off for
                                     other aligners
                                     Default: true
        *     --outputHDFSDir        HDFS Directory to store the variant calls
                                     in
              --resolution           Size of the bins to tile the genome with
                                     Default: 25
        *     --targetIsize          Mean insert size of the library
                                     Default: 0
        *     --targetIsizeSD        Standard deviation of the insert size of
                                     the library
                                     Default: 0
              --threshold            Likelihood ratio threshold to call a
                                     variant
                                     Default: 1.68

    copyToS3      Upload a file to Amazon S3 using multi-part upload
      Usage: copyToS3 [options]
  Options:
        *     --S3Bucket   S3 Bucket to upload to
        *     --fileName   Path to the file to be uploaded on the local
                           filesystem

    launchCluster      Use whirr to create a new cluster in the cloud using 
                       whirr/cloudbreak-whirr.properties
      Usage: launchCluster [options]

    runScriptOnCluster      Execute a script on one node of the currently running 
                            cloud cluster
      Usage: runScriptOnCluster [options]
  Options:
        *     --fileName   Path on the local filesystem of the script to run

    destroyCluster      Destroy the currently running whirr cluster
      Usage: destroyCluster [options]

    summarizeAlignments      Gather statistics about a set of alignments: number of reads, 
                             number of mappings, and total number of mismatches
      Usage: summarizeAlignments [options]
  Options:
              --aligner        Format of the alignment records
                               (sam|mrfast|novoalign)
                               Default: sam
        *     --inputHDFSDir   HDFS path of the directory that holds the
                               alignments

    exportGMMResults      Export wig files that contain the GMM features across 
                          the entire genome
      Usage: exportGMMResults [options]
  Options:
        *     --faidx          Local path to the chromosome length file
        *     --inputHDFSDir   HDFS path to the directory holding the GMM
                               features
        *     --outputPrefix   Prefix of the names of the files to create
              --resolution     Bin size that the GMM features were computed for
                               Default: 25

    dumpReadsWithScores      Dump all read pairs that span the given region with their 
                             deletion scores to BED format (debugging)
      Usage: dumpReadsWithScores [options]
  Options:
              --aligner                            Format of the alignment
                                                   records (sam|mrfast|novoalign)
                                                   Default: sam
        *     --inputFileDescriptor                HDFS path to the directory
                                                   that holds the alignment records
              --maxInsertSize                      Maximum possible insert size
                                                   to consider
                                                   Default: 500000
              --minScore                           Minimum alignment score (SAM
                                                   tag AS); all reads with lower AS
                                                   will be ignored
                                                   Default: -1
        *     --outputHDFSDir                      HDFS path to the directory
                                                   that will hold the output
        *     --region                             region to find read pairs
                                                   for, in chr:start-end format
              --stripChromosomeNamesAtWhitespace   Clip chromosome names from
                                                   the reference at the first
                                                   whitespace so they match with alignment
                                                   fields
                                                   Default: false

    debugReadPairInfo      Compute the raw data that goes into the GMM fit procedure for 
                           each bin (use with filter to debug a particular locus)
      Usage: debugReadPairInfo [options]
  Options:
              --aligner                 Format of the alignment records
                                        (sam|mrfast|novoalign)
                                        Default: sam
        *     --chrFilter               Print info for alignments in the region
                                        chrFilter:startFilter-endFilter
        *     --endFilter               see chrFilter
              --excludePairsMappingIn   HDFS path to a BED file. Any reads
                                        mapped within those intervals will be excluded
                                        from the processing
        *     --faidx                   HDFS path to the chromosome length file
                                        for the reference genome
        *     --inputFileDescriptor     HDFS path to the directory that holds
                                        the alignment records
              --mapabilityWeighting     HDFS path to a BigWig file containing
                                        genome uniqness scores. If specified,
                                        Cloudbreak will weight reads by the uniqueness of
                                        the regions they mapped to
              --maxInsertSize           Maximum insert size to consider (= max
                                        size of deletion detectable)
                                        Default: 500000
              --minScore                Minimum alignment score (SAM tag AS);
                                        all reads with lower AS will be ignored
                                        Default: -1
        *     --outputHDFSDir           HDFS directory to hold the output
              --resolution              Size of the bins to tile the genome with
                                        Default: 25
        *     --startFilter             see chrFilter

    findAlignment      Find an alignment record that matches the input string
      Usage: findAlignment [options]
  Options:
        *     --HDFSAlignmentsDir   HDFS path to the directory that stores the
                                    alignment data
        *     --outputHDFSDir       HDFS path to the directory in which to put
                                    the results
        *     --read                Read name or portion of the read name to
                                    search for

    sortGMMResults      Sort and merge GMM Results (use with one reducer to get all 
                        GMM feature results into a single file
      Usage: sortGMMResults [options]
  Options:
        *     --inputHDFSDir    HDFS path to the directory holding the GMM
                                features
        *     --outputHDFSDir   Directory in which to put the results
\end{verbatim}

\end{document}